\documentclass[aps,prd,reprint,superscriptaddress,raggedbottom,amsmath,amssymb,floatfix]{revtex4-2}
\usepackage{graphicx}
\usepackage{bm}
\usepackage{booktabs}
\usepackage[unicode,colorlinks=true,linkcolor=blue,citecolor=blue,urlcolor=blue]{hyperref}
\numberwithin{equation}{section}
\renewcommand{\theequation}{\arabic{section}.\arabic{equation}}

\makeatletter
\newenvironment{inlinefigure}{%
  \par\addvspace{4pt}\noindent
  \begin{minipage}{\linewidth}\def\@captype{figure}%
}{%
  \end{minipage}\par\addvspace{4pt}%
}
\makeatother
\newenvironment{widefigure}{%
  \par\onecolumngrid\begin{inlinefigure}%
}{%
  \end{inlinefigure}\twocolumngrid
}

\begin{document}

\title{Thermal Synchrotron Flares Driven by Magnetic Reconnection in the Plunging Region of a Kerr Black Hole}
\author{Fan Zhou}
\affiliation{School of Physics and Astronomy, Beijing Normal University, Beijing 100875, China}
\author{Bin Chen}
\email{chenbin1@nbu.edu.cn}
\affiliation{Institute of Fundamental Physics and Quantum Technology, \& School of Physical Science and Technology, Ningbo University, Ningbo, Zhejiang 315211, China}
\affiliation{Zhejiang Key Laboratory of Extreme Universe, \& BINGO Center, Ningbo University, Ningbo, Zhejiang 315211, China }
\author{Minyong Guo}
\email{minyongguo@bnu.edu.cn}
\affiliation{School of Physics and Astronomy, Beijing Normal University, Beijing 100875, China}
\affiliation{Key Laboratory of Multiscale Spin Physics (Beijing Normal University), Ministry of Education, Beijing 100875, China}

\begin{abstract}
Magnetic reconnection is a promising energy source for black hole flares. Ideal magnetohydrodynamic simulations may underestimate the energy release rate of collisionless reconnection, while changes in the global flow and magnetic field make its radiative contribution difficult to isolate. To study the effects of this heating channel on black hole images and time-dependent signals, we construct a semianalytical thick-disk model that tracks how energy from reconnection outflows changes the disk temperature and emission in a fixed background flow. For a Kerr black hole with the mass and distance of M87$^*$, the quiescent $230\,\mathrm{GHz}$ image shows a ring that is brighter on one side, as seen by the Event Horizon Telescope. During sustained heating, arcs of the ring brighten markedly and the light curves develop multiple peaks. The peak total flux exceeds its quiescent value by approximately $19\%$ in fast light and $25\%$ in slow light. The thermal response of Faraday rotation reduces polarization cancellation along the line of sight, increasing the image-averaged linear polarization fraction. These results suggest that localized magnetic reconnection could power transient millimeter brightening in the ring of M87$^*$.
\end{abstract}
\maketitle

\section{Introduction}
\label{sec:introduction}

The Event Horizon Telescope (EHT) images of M87$^*$ and Sgr~A$^*$ have enabled direct studies of the emitting structures around supermassive black holes on horizon scales \cite{EHTM87I,EHTM87II,EHTM87III,EHTM87IV,EHTSgrAI,EHTSgrAII,EHTSgrAIII}. Analyses of the bright ring, black hole shadow, and emission models provide observational constraints on photon propagation and accretion physics in strong gravitational fields \cite{R17,EHTM87VI,EHTSgrAIV,EHTSgrAV,EHTSgrAVI}. Linear and circular polarization measurements further constrain the near-horizon magnetic field, electron thermodynamics, and Faraday effects \cite{EHTM87VII,EHTM87VIII,EHTM87IX,EHTSgrAVII,EHTSgrAVIII}. Observations across multiple epochs show that a relatively stable ring size can coexist with evolving brightness and polarization structures \cite{EHTM87PersistentI,EHTM87PersistentII,EHTM87Variability2025}. Interpreting these images and their variability requires connecting spacetime geometry, plasma dynamics, and radiative processes \cite{Zhang2024Thick,Hou2025NearHorizonPol}.

Variations in brightness and polarization during black hole flares provide clues to localized energy release and plasma motion. Astrometric and polarimetric observations of near-infrared flares in Sgr~A$^*$ have provided evidence for hot spot motion near the innermost stable circular orbit (ISCO) \cite{GRAVITY2018,GRAVITY2023}. Localized hot spot models connect the motion of emitting regions with relativistic propagation effects to investigate light curves, polarization patterns, and temporal correlations \cite{R07,R08,Zeng2026KerrAdS,Wang2026KerrSen,Chen2025HotspotPol,Zhang2025Autocorrelation}. Explaining the physical origin of these signals also requires linking the motion and emission of the plasma to specific energy release and thermal evolution processes.

Magnetic reconnection is a promising energy source for black hole flares. It converts magnetic energy into bulk kinetic energy, heat, and nonthermal particle energy. In strongly magnetized plasmas, reconnection outflows can reach relativistic speeds and transport the released energy into the surrounding medium \cite{Lyubarsky2005}. Comisso and Asenjo studied relativistic magnetic reconnection in curved spacetime and proposed a mechanism for extracting black hole spin energy through magnetic reconnection in the ergosphere \cite{Asenjo2017Kerr,Comisso2018Collisionless,Asenjo2019Gravity,R04}. These processes can form localized heated structures near the black hole \cite{R05,R06,ElMellah2022Corona}.

The macroscopic evolution of these processes in accretion flows can be studied using general relativistic magnetohydrodynamic (GRMHD) simulations. Existing simulations have revealed near-horizon current sheets, heated plasma structures, and magnetic flux ejections \cite{Nathanail2020Plasmoids,Nathanail2022Reconnection,R06}. These calculations usually adopt ideal MHD, in which reconnection relies on numerical dissipation. The reconnection rate is commonly measured by the ratio of the inflow speed into a local current sheet to the upstream Alfv\'en speed. In high-resolution ideal GRMHD simulations, plasmoid-mediated reconnection can converge to a rate of approximately $0.01$ \cite{R06}, whereas collisionless reconnection in strongly magnetized plasmas with weak guide fields can reach approximately $0.1$ \cite{Bransgrove2021MagneticHair,Galishnikova2023Collisionless}. For comparable upstream fields and current sheet scales, this difference suggests that ideal GRMHD may underestimate the rate of magnetic energy conversion into plasma energy, thereby affecting the predicted radiative variability. General relativistic particle-in-cell (GRPIC) simulations can describe collisionless reconnection and particle acceleration self-consistently, but resolving both plasma microscales and black hole scales is computationally expensive. Existing global accretion comparisons have therefore used two-dimensional axisymmetric configurations \cite{Galishnikova2023Collisionless}.

A further difficulty is isolating the contribution of reconnection heating to the radiation from global simulations. The partition of energy between electrons and ions depends on the upstream plasma state \cite{R19}, while the treatment of electron heating and radiative cooling changes the images and their variability \cite{R15,Jiang2023TwoTemperature,CruzOsorio2026ElectronTemperature,Long2026CoolingPolarization}. The electron distribution also affects flux and polarization through emission, absorption, and Faraday effects \cite{R09}. During magnetic flux eruptions, reconnection heating is coupled to turbulence, density evacuation, and changes in the flow and magnetic field. For example, different electron temperature models can produce either brightening or dimming at $230\,\mathrm{GHz}$, as changes in the emission of heated plasma compete with evacuation of the inner flow \cite{Jia2023MillimeterFlares}. The net variability during an eruption therefore cannot be attributed directly to reconnection heating alone.

Analytical and semianalytical models can isolate the radiative role of reconnection heating by prescribing the energy input and background conditions. Previous models have studied reconnection rates and outflow energies in Kerr spacetime and extended reconnection-driven extraction of black hole spin energy to plunging plasma inside the ISCO \cite{Fan2024Fast,Fan2025Forces,R04,Chen2024Plunging,Shen:2024sdr,Wan:2026aio}. Reconnection-heated corona models have also connected magnetic energy release with coronal conduction and cooling \cite{Liu2002Corona}. More recently, Li incorporated the energy and angular momentum fluxes of reconnection outflows into the conservation equations of a stationary thin disk and obtained the radial response of its surface emission \cite{R01}. Extending such stationary disk responses to time-dependent off-equatorial imaging of localized reconnection events requires following the energy transferred by outflows along their paths, the delay in energy transport from the equatorial layer to the off-equatorial regions, and the subsequent evolution of the heated plasma, together with radiative transfer.

In this work, we use a semianalytical thick-disk model to investigate whether localized reconnection inside the ISCO can produce pronounced thermal synchrotron flares by transferring outflow energy to the equatorial layer and then to the off-equatorial regions. To isolate this heating channel, we prescribe the outflow power and keep the background density, velocity, magnetic field, and disk geometry fixed. Reconnection heating is the only additional energy source, and we evolve the resulting internal energy perturbation.

We adopt the mass and distance scales of M87$^*$ in Kerr spacetime and use our previously developed CoportSL polarized radiative-transfer framework~\cite{Huang2024Coport,R09,CoportSL} to calculate $230\,\mathrm{GHz}$ light curves and polarized images, comparing fast and slow light. Decomposing the flux by fixed emitting region identifies its main origins. We then track their azimuthal evolution, foreground optical depths, and arrival-time distributions to interpret successive peaks during sustained heating. Control calculations of the Faraday rotation coefficient examine the increase in the image-averaged linear polarization fraction near the second flux peak.

The remainder of this paper is organized as follows. In Section~\ref{sec:model}, we introduce the background thick-disk model. In Section~\ref{sec:heating}, we construct the model for reconnection outflow power, energy transport from the equatorial layer to both off-equatorial regions, and thermal perturbation evolution. In Section~\ref{sec:kerr}, we present the Kerr configuration and polarized imaging method. In Section~\ref{sec:results}, we analyze the time-dependent images, light curves, and polarization. In Section~\ref{sec:discussion}, we summarize the main results and discuss the model's applicability. We adopt the metric signature $(-,+,+,+)$ and use $G=M=c=1$ for geometric quantities, with lengths and times measured in $r_g=GM/c^2$ and $t_g=GM/c^3$, respectively. Thermodynamic and radiative quantities are expressed in CGS units, with Gaussian electromagnetic conventions.

\section{Thick-Disk Model}
\label{sec:model}

\subsection{Spacetime and Thick-Disk Geometry}

We consider a stationary, axisymmetric, circular black hole spacetime with equatorial reflection symmetry. Outside the horizon, we choose reference coordinates $x^\mu=(t,r,\theta,\phi)$ adapted to time translations and axial rotations, with orthogonal meridional directions, so that the metric takes the form
\begin{equation}
\begin{aligned}
ds^2={}&g_{tt}dt^2+2g_{t\phi}\,dt\,d\phi\\
&+g_{rr}dr^2+g_{\theta\theta}d\theta^2+g_{\phi\phi}d\phi^2.
\end{aligned}
\label{eq:metric}
\end{equation}
All metric components depend only on $r,\theta$. Here $r$ labels the chosen radial surfaces, and $\theta$ is the polar angle measured from the rotation axis, with the equatorial plane at $\theta=\pi/2$. The background distributions and heating model are defined in the chosen reference coordinates. The event horizon radius and the ISCO radius of the selected orbital branch are denoted by $r_H$ and $r_I$, respectively.

We adopt a magnetized thick-disk model, divided into a dense equatorial layer and hot, more tenuous magnetized regions above and below it. The equatorial layer extends inside the ISCO \cite{R25,R26}. The thick-disk background extends from outside the horizon to $r_o$ over the fixed angular domain $\theta_f\leq\theta\leq\pi-\theta_f$. The visible emitting region excludes the equatorial layer superposed on this background. The angular boundary sets the vertical extent of the thick disk; the polar funnel contains no radiating plasma.

At fixed $r$, points within a proper polar distance $H_{\rm eq}(r)\ll r$ of the equatorial plane form the equatorial layer. This distance is measured with the line element $\sqrt{g_{\theta\theta}}\,d\theta$. Inside the layer, we use a frequency-independent comoving gray absorption coefficient $\alpha_{\rm eq}=\tau_{\rm eq}/(2r_gH_{\rm eq})$, where $H_{\rm eq}$ is measured in units of $r_g$ and $\tau_{\rm eq}\gg1$ is the reference optical depth through the full vertical thickness. The thickness and optical depth are prescribed independently; the large optical depth makes the layer obscure radiation from the opposite off-equatorial region at the millimeter frequencies considered here.

Thermal synchrotron emission from electrons in the off-equatorial regions supplies the millimeter images \cite{R22}. Emission within the equatorial layer is set to zero, retaining only gray absorption and obscuration. Unless explicitly labeled, the density, velocity, magnetic field, and thermodynamic variables below describe the off-equatorial regions. The reference equatorial four-velocity $u_{\rm eq}^\mu$ follows equatorial circular geodesics outside the ISCO and inward geodesics inheriting the ISCO circular-orbit energy and angular momentum inside it. We extend the equatorial four-velocity at the same radius into the layer and renormalize it using the local metric.

\subsection{Off-Equatorial Density, Pressure, and Motion}
\label{sec:background}

We prescribe a stationary, axisymmetric rotating inflow in the off-equatorial regions, with $u^\theta=0$ and $u^r<0$, so that streamlines follow surfaces of fixed polar angle. We assume a pure hydrogen plasma, with electron number density $n_e=\rho/m_p$, where $\rho$ is the rest-mass density and $m_p$ is the proton mass. Motivated by the radial power-law parameterization of density and temperature in coronal radio models \cite{R22}, we extend the density and pressure power laws throughout the thick-disk background outside the horizon:
\begin{equation}
\begin{aligned}
n_e(r)&=n_*\left(\frac r{r_*}\right)^{-q},\\
p_0(r)&=p_*\left(\frac r{r_*}\right)^{-s}.
\end{aligned}
\label{eq:background}
\end{equation}
The subscript 0 denotes the quiescent state. The parameters $n_*,p_*$ are the electron number density and gas pressure at the reference radius $r_*$, while $q,s$ are radial indices prescribed in this work. The corresponding temperature scaling is $p_0/n_e\propto r^{q-s}$. The gas internal energy density is $u_{\text{g},0}=p_0/(\Gamma-1)$, where $\Gamma$ is the adiabatic index.

For $r>r_I$, equatorial circular geodesics provide the rotation reference, with four-velocity \cite{R02,R03}
\begin{equation}
u_{\rm eq}^\mu(r>r_I,\pi/2)=u_K^t(1,0,0,\Omega_K).
\label{eq:circular}
\end{equation}
Here $\Omega_K$ is the angular velocity of an equatorial circular geodesic, determined by $\partial_r g_{tt}+2\Omega_K\partial_r g_{t\phi}+\Omega_K^2\partial_r g_{\phi\phi}=0$, and $u_K^t=(-g_{tt}-2\Omega_Kg_{t\phi}-\Omega_K^2g_{\phi\phi})^{-1/2}$ is the corresponding time component, fixed by four-velocity normalization.

The radial motion follows from mass conservation on both sides of $r_I$. The continuity equation $\nabla_\mu(\rho u^\mu)=0$ gives \cite{Gammie2003HARM}
\begin{equation}
\sqrt{-g}\,\rho u^r=F(\theta),
\label{eq:continuity}
\end{equation}
where $g=\det g_{\mu\nu}$ and $F(\theta)<0$ is the mass flux factor conserved along each streamline. Denoting the background mass accretion rate by $\dot M>0$, its integral over the thick-disk angular domain is
\begin{equation}
\dot M=-2\pi c r_g^2\int_{\theta_f}^{\pi-\theta_f}F(\theta)\,d\theta.
\label{eq:accretionrate}
\end{equation}
Here $\rho$ and $\dot M$ are in CGS units, and $c r_g^2$ converts the geometrical velocity and area to a physical mass flow rate. The equatorial layer is superposed as an effective absorber and energy-transfer layer; its additional mass flux is neglected, and obscuration does not alter the definition of the background flux. Assuming equal mass flow per unit solid angle gives
\begin{equation}
F(\theta)=-\frac{\dot M\sin\theta}{4\pi c r_g^2\cos\theta_f}.
\label{eq:radialvelocity}
\end{equation}
For the prescribed power-law density and $\dot M$, Equation~\eqref{eq:continuity} determines the radial velocity throughout the background.

Motivated by magnetic coupling between the off-equatorial plasma and the equatorial layer, we approximately assign them the same angular velocity at the same radius for $r>r_I$:
\begin{equation}
u^\phi(r>r_I,\theta)=\Omega_K(r,\pi/2)u^t.
\label{eq:outervelocity}
\end{equation}
The time component is fixed by $u_\mu u^\mu=-1$, with $u^t>0$. The specific angular momentum $u_\phi$ of this outer velocity field at the ISCO boundary of each streamline is denoted by
\begin{equation}
L_I(\theta)=\left[(g_{t\phi}+\Omega_Kg_{\phi\phi})u^t\right]_{(r_I,\theta)}.
\label{eq:iscoangularmomentum}
\end{equation}
For $r<r_I$, we adopt conservation of specific angular momentum along each streamline:
\begin{equation}
u_\phi(r<r_I,\theta)=L_I(\theta).
\label{eq:inflow}
\end{equation}
The radial velocity still follows from Equation~\eqref{eq:continuity}. The azimuthal condition and four-velocity normalization determine $u^t,u^\phi$, selecting the future-directed timelike solution. Inheriting the ISCO angular momentum at the same polar angle makes the four-velocity continuous across $r_I$. The density and pressure follow Equation~\eqref{eq:background} throughout the background.

\subsection{Magnetic Field and Electron Temperature}
\label{sec:field}

We adopt an ordered magnetic field with toroidal and poloidal components, expressed through the magnetic flux variables $\mathcal B^i$ commonly used in ideal GRMHD \cite{Gammie2003HARM}. The poloidal field is constructed from a flux function $\Psi(r,\theta)$, and the toroidal component is an axisymmetric function $\mathcal B^\phi(r,\theta)$:
\begin{equation}
(\mathcal B^r,\mathcal B^\theta,\mathcal B^\phi)
=\left(\frac{\partial_\theta\Psi}{\sqrt{-g}},
-\frac{\partial_r\Psi}{\sqrt{-g}},\mathcal B^\phi(r,\theta)\right).
\label{eq:flux}
\end{equation}
This construction satisfies $\partial_i(\sqrt{-g}\mathcal B^i)=0$.

For a given off-equatorial four-velocity, the comoving magnetic field four-vector $b^\mu$ follows from the standard transformation \cite{Gammie2003HARM},
\begin{equation}
b^t=\mathcal B^iu_i,\qquad
b^i=\frac{\mathcal B^i+b^tu^i}{u^t}.
\label{eq:bfield}
\end{equation}
The comoving field obeys $u_\mu b^\mu=0$. We write its squared norm as $b^2\equiv b^\mu b_\mu$, with field strength $B=\sqrt{b^2}$. Both the plasma $\beta$ and synchrotron radiation are calculated using the resulting local comoving field.

The electron temperature is determined jointly by the total gas pressure and the ion-to-electron temperature ratio. For the latter, we adopt the prescription introduced by Mo\'{s}cibrodzka et al.\ and widely used in black hole imaging \cite{Moscibrodzka2016Temperature,R17},
\begin{equation}
\frac{T_i}{T_e}=
\frac{R_{\rm low}+R_{\rm high}\beta^2}
{1+\beta^2},\qquad
p=n_ek_B(T_i+T_e).
\label{eq:temperature}
\end{equation}
Here $\beta=8\pi p/b^2$ is the ratio of gas to magnetic pressure, and $R_{\rm low},R_{\rm high}$ give the ion-to-electron temperature ratios in the magnetically and gas-pressure dominated limits, respectively. The quiescent and heated states use the same composition and temperature prescription.

\section{Reconnection-driven Heating}
\label{sec:heating}

\subsection{Near-horizon Reconnection Source and Relativistic Outflow}
\label{sec:deposition}

High-resolution simulations have revealed strong reconnection and heated structures in low-density, highly magnetized regions near the horizon \cite{R05,R06}. Motivated by this picture, we select a localized reconnection source with initial radius $r_0$ satisfying $r_H<r_0<r_I$ and follow the energy transferred by its outflow through the equatorial layer to the off-equatorial regions.

We represent the localized reconnection source by the center of the current sheet, which follows an inward equatorial geodesic inheriting the specific energy and angular momentum of the ISCO circular orbit \cite{R03,Chen2024Plunging}. These constants follow from the reference circular orbit in Equation~\eqref{eq:circular}:
\begin{align}
E_K(r_I,\pi/2)&=-\left[(g_{tt}+\Omega_Kg_{t\phi})u_K^t\right]_{(r_I,\pi/2)},\nonumber\\
L_K(r_I,\pi/2)&=\left[(g_{t\phi}+\Omega_Kg_{\phi\phi})u_K^t\right]_{(r_I,\pi/2)}.
\label{eq:iscoconstants}
\end{align}
The source four-velocity $(u_{\rm s})^\mu$ satisfies $(u_{\rm s})_t=-E_K(r_I,\pi/2)$, $(u_{\rm s})_\phi=L_K(r_I,\pi/2)$, $(u_{\rm s})^\theta=0$, and $(u_{\rm s})_\mu(u_{\rm s})^\mu=-1$, selecting the future-directed inward branch.

The local upstream plasma is taken to be comoving with the source, and the launch time is denoted by $t_s$. In the source rest frame, the magnetization is $\sigma=B_0^2/(4\pi w_0)$, where $B_0$ is the upstream magnetic field strength and $w_0$ is the matter enthalpy density including rest-mass energy. We estimate the outflow speed using the relativistic Alfv\'en speed and prescribe its launch direction \cite{R04,Chen2024Plunging}:
\begin{equation}
\begin{gathered}
v_{\rm out}=\sqrt{\frac{\sigma}{1+\sigma}},\\
u_b^\mu=\gamma_{\rm out}
\left[(u_{\rm s})^\mu+v_{\rm out}\left(\sin\xi\,e_{(r)}^\mu+\cos\xi\,e_{(\phi)}^\mu\right)\right].
\end{gathered}
\label{eq:outflow}
\end{equation}
Here $\gamma_{\rm out}=\sqrt{1+\sigma}$ is the outflow Lorentz factor in the source rest frame. The orthonormal basis vectors $e_{(r)}^\mu,e_{(\phi)}^\mu$ are constructed in that frame. We first project the axial Killing vector $\partial_\phi$ into the rest space orthogonal to $(u_{\rm s})^\mu$ and normalize it to obtain $e_{(\phi)}^\mu$. We then orthogonalize $\partial_r$ against $(u_{\rm s})^\mu$ and $e_{(\phi)}^\mu$, normalize the result, and choose $e_{(r)}^r>0$. The angle $\xi$ is measured from the positive azimuthal direction.

For each launch, this four-velocity determines the specific Killing energy $-u_{bt}$ and angular momentum $u_{b\phi}$. We integrate the corresponding equatorial timelike geodesic to obtain the outflow center and its propagation time. As the source plunges inward, its position and four-velocity change, so continuous launching produces a family of timelike geodesics parameterized by launch time. At a fixed coordinate time, the centers launched at different times form an evolving curve whose shape reflects both the source motion and the subsequent outflow propagation.

For positive-energy outflows along these trajectories, we define the energy using the time-translation Killing field normalized at infinity and prescribe a constant power per unit coordinate time,
\begin{equation}
L_0=\epsilon_b\dot Mc^2.
\label{eq:power}
\end{equation}
Here $\dot M$ is the same background mass accretion rate defined in Equation~\eqref{eq:accretionrate}, and $\epsilon_b$ is the ratio of outflow power to the accretion rest-mass energy flux $\dot Mc^2$. The first launch occurs at $t_s=0$. Injection ceases at $t_{\rm end}$, when the source reaches a prescribed cutoff radius outside the horizon, while previously launched outflows continue to propagate.

For fixed outflow kinematics and integration geometry, Ref.~\cite{R01} sets the energy-input amplitude through the outflow enthalpy density, whereas we directly normalize the integrated Killing power with $\epsilon_b$. Both choices retain one independent power amplitude parameter.

The reference scale of $\epsilon_b$ can be estimated from the magnetic energy entering the reconnection layer. For a strongly magnetized upstream, the Alfv\'en speed approaches $c$. Taking a collisionless reconnection rate $v_{\rm in}/v_A\simeq0.1$ \cite{Bransgrove2021MagneticHair,Galishnikova2023Collisionless}, the local Poynting power entering the current sheet is approximately
\begin{equation}
P_{\rm rec}^{\rm loc}\simeq\frac{B_{\rm rec}^2}{4\pi}v_{\rm in}A_{\rm in},
\label{eq:magneticbudget}
\end{equation}
where $B_{\rm rec}$ is the upstream magnetic field component that reverses across the sheet and $A_{\rm in}$ is the total area of its two inflow faces. Reference~\cite{R06} estimates a near-horizon field of $20$--$200\,\mathrm G$ for M87$^*$ and adopts $100\,\mathrm G$ as a fiducial upstream value. For $M=6.5\times10^9M_\odot$ and dimensionless Kerr spin $a=0.9375$, if the full equatorial annulus between the horizon and the ISCO reconnects simultaneously, a flat-area approximation gives $A_{\rm in}\simeq2\pi(r_I^2-r_H^2)\simeq15r_g^2$. With $B_{\rm rec}=100\,\mathrm G$ and $v_{\rm in}=0.1c$, Eq.~\eqref{eq:magneticbudget} yields $P_{\rm rec}^{\rm loc}\simeq3\times10^{43}\,\mathrm{erg\,s^{-1}}$, about $0.6$ of the rest-mass energy flux for $\dot M=10^{-3}M_\odot\,\mathrm{yr^{-1}}$. This budget motivates $\epsilon_b\sim0.1$--$1$ as a reference scale for efficient reconnection over an extended region. The estimate describes the total local energy supply; conversion to the Killing power of the single outflow in Eq.~\eqref{eq:power} also depends on energy partition and relativistic transformations. More localized reconnection, weaker reconnecting fields, or lower conversion efficiencies can yield smaller ratios.

\subsection{Outflow Heating of the Accretion Disk}

The reconnection outflow propagates in the equatorial plane and gradually transfers energy to the equatorial layer through interaction with equatorial-layer material. For each launch, we accumulate the arc length $s$ from the launch point along the spatial projection of the central worldline on constant-$t$ slices, using the equatorial spatial line element $ds^2=g_{rr}dr^2+g_{\phi\phi}d\phi^2$. The remaining energy fraction relative to the initial Killing energy is prescribed as
\begin{equation}
f(s)=\left(1+\frac{s}{\ell}\right)^{-m}.
\label{eq:loss}
\end{equation}
The parameter $\ell$ sets the loss length, and $m$ controls the decay. The Killing energy transferred to each path segment is the initial energy multiplied by the decrease in $f$ along that segment.

Summing the loss rates from all launches gives the power transferred to the equatorial layer at each time. To account for the finite outflow width, we broaden the instantaneous central curve in its orthogonal direction within the equatorial plane at fixed coordinate time $t$. Let $s_\perp$ be the signed proper distance in this direction. We adopt a Gaussian weight $w(s_\perp)\propto\exp[-s_\perp^2/(2b_\perp^2)]$, truncated at $[-6b_\perp,6b_\perp]$ and normalized, where $b_\perp$ is the width parameter. Denoting by $Y(t_s,s_\perp;t)$ the position offset by $s_\perp$ from the corresponding curve point, the spatial distribution of the transferred power is
\begin{equation}
\begin{aligned}
\mathcal D(R,\phi,t)
={}&\int dt_s\,ds_\perp\,h(t_s,t)w(s_\perp)\\
&\times\delta^{(2)}\!\left[(R,\phi)-Y(t_s,s_\perp;t)\right].
\end{aligned}
\label{eq:deposit}
\end{equation}
Here $h(t_s,t)=-df[s(t_s,t)]/dt$ is the fraction of initial energy lost per unit coordinate time by the outflow launched at $t_s$. The Killing power received by an equatorial area element is $dP_{\rm dep}^{\infty}=L_0\mathcal D(R,\phi,t)\,dR\,d\phi$. The integral includes only launches with $0\leq t_s\leq\min(t,t_{\rm end})$ whose centers have not reached a tracking boundary; otherwise $h=0$. The inner tracking radius equals the source injection cutoff radius, and the outer radius is $r_{\rm eq}$. Any energy remaining when a trajectory terminates is no longer transferred to the equatorial layer. On the equatorial plane, $R=r$. The two-dimensional $\delta$ function is defined with respect to the coordinate area $dR\,d\phi$, and only the broadened distribution within $r_H<R<r_{\rm eq}$ is retained.

After the equatorial layer receives outflow energy and is heated, part of the thermal energy propagates into the regions above and below it. We use an effective prescription for this internal energy transport. The transport model specifies the time-dependent heat input at each fixed position, and the arriving energy heats the fluid element occupying that position at that time. Fluid elements move with the background velocity field and advect the excess internal energy they acquire.

Using $R=r\sin\theta$ and $z=r\cos\theta$, we construct transport channels that extend from the equatorial plane $z=0$ into the off-equatorial regions on both sides on cylinders of fixed $R$, allowing their azimuth to change during propagation. The effective transport speed is set to the local speed of light, representing the upper limit of rapid energy transport. The channel tangent satisfies $g_{\mu\nu}v^\mu v^\nu=0$ and $v_\phi=0$. Above the equatorial layer, at $z\geq0$, let $t_v(R,z)$ be the coordinate travel time from the equatorial layer to $(R,z)$ and $\phi_v(R,z)$ the azimuthal shift accumulated during propagation. These constraints and Equation~\eqref{eq:metric} give
\begin{equation}
\begin{aligned}
\frac{\partial t_v}{\partial z}
&=\sqrt{\frac{(z^2/r^2)g_{rr}+(R^2/r^4)g_{\theta\theta}}
{-g_{tt}+g_{t\phi}^{\,2}/g_{\phi\phi}}},\\
\frac{\partial\phi_v}{\partial z}
&=-\frac{g_{t\phi}}{g_{\phi\phi}}\frac{\partial t_v}{\partial z}.
\end{aligned}
\label{eq:channel}
\end{equation}
All metric components are evaluated in the original reference coordinates at $r=\sqrt{R^2+z^2}$ and $\cos\theta=z/r$. We integrate where $g_{\phi\phi}>0$ and $-g_{tt}+g_{t\phi}^{\,2}/g_{\phi\phi}>0$, taking the positive root for future-directed propagation toward increasing $z$. The initial conditions are $t_v(R,0)=\phi_v(R,0)=0$, and the channels below the equatorial layer follow by equatorial reflection symmetry.

A fraction $\eta$ of the energy received by the equatorial layer is thermalized, and a fraction $f_{\rm off}$ of that heat is transferred to the off-equatorial regions on both sides. Within each two-sided column at fixed $R$, all heights share a common heating normalization, while the local input accounts for the propagation delay and azimuthal shift. The heating rate $q_{\rm off}$ per unit comoving volume and proper time is
\begin{equation}
q_{\rm off}(t,r,\theta,\phi)
=\frac{\eta f_{\rm off}L_0}{r_g^3C(R)}
\mathcal D\!\left[R,\phi-\phi_v,t-t_v\right].
\label{eq:heating}
\end{equation}
The shifts in time and azimuth mean that the heat input at the off-equatorial position $(R,\phi,z)$ and time $t$ is supplied by the outflow power received at the equatorial position $(R,\phi-\phi_v)$ at time $t-t_v$. The normalization factor $C(R)=\int_{\rm column}(-u_t)(\sqrt{-g}/r)\,dz$ is integrated over both off-equatorial regions at fixed $R$, bounded by the equatorial-layer surfaces, funnel boundary, and outer radius $r_o$. The factor $-u_t$ converts comoving heat input to Killing power, while $\sqrt{-g}/r$ is the integration measure obtained by transforming $(r,\theta)$ to $(R,z)$. Integrated over the full heating history and off-equatorial regions, the Killing energy supplied as heat is $\eta f_{\rm off}$ times the energy received by the equatorial layer. Propagation delays change the instantaneous arrival power but leave this total unchanged. The heating rate is set to zero when the corresponding equatorial-layer time satisfies $t-t_v<0$.

\subsection{Advection, Adiabatic Evolution, and Cooling}
\label{sec:thermal}

We denote the heating rate per unit comoving volume and proper time by $q_{\rm heat}$ and describe cooling phenomenologically as $u_{\text{g}}/(t_g\tau_c)$, where $\tau_c$ is the dimensionless proper cooling time normalized by $t_g$. Introducing these source terms into the stress-energy equation, projecting along the fluid four-velocity, and using mass conservation to eliminate the rest-mass energy contribution gives the gas internal energy equation \cite{R15},
\begin{equation}
\nabla_\mu(u_{\text{g}}u^\mu)+p\nabla_\mu u^\mu
=t_gq_{\rm heat}-\frac{u_{\text{g}}}{\tau_c}.
\label{eq:thermalbalance}
\end{equation}
The quiescent heating rate is $q_{{\rm heat},0}$, while the heated state has $q_{\rm heat}=q_{{\rm heat},0}+q_{\rm off}$. Both states use the same density, velocity, and magnetic field background, cooling time, and constant-$\Gamma$ equation of state $p=(\Gamma-1)u_{\text{g}}$. Defining the excess internal energy density as $\delta u_{\text{g}}=u_{\text{g}}-u_{\text{g},0}$, subtracting the quiescent equation from its heated counterpart, and expanding the divergence gives
\begin{equation}
u^\mu\partial_\mu\delta u_{\text{g}}+
\Gamma(\nabla_\mu u^\mu)\delta u_{\text{g}}
=t_gq_{\rm off}-\frac{\delta u_{\text{g}}}{\tau_c}.
\label{eq:thermal}
\end{equation}
The background satisfies $u^\theta=0$, so $u^\mu\partial_\mu\delta u_{\text{g}}=u^t\partial_t\delta u_{\text{g}}+u^r\partial_r\delta u_{\text{g}}+u^\phi\partial_\phi\delta u_{\text{g}}$, where the radial and azimuthal terms describe advection. The second term on the left combines volume changes and pressure work, describing adiabatic compression or expansion.

We set the excess internal energy to zero for $t\leq0$ and at the inflow boundary $r=r_o$. Using mass conservation, $u^\mu\partial_\mu\rho=-\rho\nabla_\mu u^\mu$, we rewrite Equation~\eqref{eq:thermal} as an evolution equation for $\delta u_{\text{g}}/\rho^\Gamma$ and integrate along a fluid trajectory to obtain
\begin{equation}
\begin{aligned}
\delta u_{\text{g}}(t,\boldsymbol x)
={}&\rho(\boldsymbol x)^\Gamma\int_{\rm history}
\frac{t_gq_{\rm off}(t',\boldsymbol x')}{\rho(\boldsymbol x')^\Gamma}\\
&\times\exp\!\left[-\frac{\tau-\tau'}{\tau_c}\right]\,d\tau'.
\end{aligned}
\label{eq:thermalintegral}
\end{equation}
Here $\tau,\tau'$ are dimensionless proper times along the fluid trajectory, normalized by $t_g$. The density ratio accounts for adiabatic compression or expansion, the exponential kernel describes cooling, and the delayed heating rate retains the history of energy transport from the equatorial layer to the off-equatorial regions. The heated gas pressure is $p=p_0+(\Gamma-1)\delta u_{\text{g}}$, and the electron temperature follows from Equation~\eqref{eq:temperature}.

\section{Source Model and Imaging in Kerr Spacetime}
\label{sec:kerr}

\begin{table*}[!tp]
\caption{Physical parameters for the adopted configuration. The reference point has proper polar height $H_*$ above the equatorial plane at radius $r_*$, and $\phi_0$ is the initial source azimuth. Angles are in radians unless explicitly marked in degrees. The gray optical depth $\tau_{\rm eq}$ is measured through the full vertical equatorial-layer thickness, $\beta_*$ sets the common field normalization scale through $B_*$, and $f_{\rm off}$ is the total fraction of thermal energy transferred to both off-equatorial regions.}
\label{tab:parameters}
\centering
\small
\begin{tabular}{lll}
\toprule
Category & Parameter & Value\\
\midrule
Spacetime/scales & $a;\ M;\ D$ & $0.9375;\ 6.5\times10^9M_\odot;\ 16.9\,\mathrm{Mpc}$\\
Reference point & $r_*;\ H_*$ & $6r_g;\ 0.3r_g$\\
Background & $n_*;\ p_*$ & $2\times10^4\,\mathrm{cm^{-3}};\ 1.8\,\mathrm{erg\,cm^{-3}}$\\
& $q;\ s;\ \Gamma$ & $2;\ 3;\ 5/3$\\
& $r_o;\ r_{\rm eq};\ \theta_f$ & $120r_g;\ 50r_g;\ \pi/4$\\
Equatorial layer & $H_{\rm eq}/r;\ \tau_{\rm eq}$ & $0.01;\ 20$\\
Magnetic field & $\beta_*;\ h_B;\ r_B;\ \zeta$ & $0.1;\ 0.09;\ 40r_g;\ \pi/4$\\
Reconnection source & $r_0;\ \phi_0;\ \sigma;\ \xi$ & $1.9r_g;\ 0.3;\ 100;\ \pi/6$\\
& Source/outflow inner cutoff & $1.01r_H$\\
Energy input & $\dot M$ & $10^{-3}M_\odot\,\mathrm{yr^{-1}}$\\
& $\epsilon_b;\ \eta;\ f_{\rm off}$ & $1;\ 0.9;\ 0.9$\\
& $b_\perp;\ \ell;\ m;\ \tau_c$ & $1.6r_g;\ 10r_g;\ 2;\ 2$\\
Electrons & $R_{\rm low};\ R_{\rm high}$ & $20;\ 100$\\
Observation & $\nu_{\rm obs};\ \theta_{\rm obs};\ \phi_{\rm obs}$ & $230\,\mathrm{GHz};\ 163^\circ;\ 10^{-4}$\\
\bottomrule
\end{tabular}
\end{table*}

\subsection{Background and Parameter Choices}
\label{sec:parameters}

In standard Boyer--Lindquist (BL) coordinates, the Kerr line element is \cite{R03}
\begin{align}
ds^2={}&-\left(1-\frac{2r}{\Sigma}\right)dt^2
-\frac{4ar\sin^2\theta}{\Sigma}\,dt\,d\phi
\nonumber\\
&+\frac{\Sigma}{\Delta}dr^2+\Sigma d\theta^2\nonumber\\
&+\left(r^2+a^2+\frac{2a^2r\sin^2\theta}{\Sigma}\right)\sin^2\theta\,d\phi^2.
\label{eq:kerr}
\end{align}
Here $\Sigma=r^2+a^2\cos^2\theta$ and $\Delta=r^2-2r+a^2$ are the usual Kerr metric functions. The prograde circular-orbit angular velocity is $\Omega_K=(r^{3/2}+a)^{-1}$, and the event horizon is at $r_H=1+\sqrt{1-a^2}$.

The source initially lies in the overlap of the plunging region and the equatorial ergosphere. The source and outflow centers use the same inner tracking cutoff outside the horizon, and outward-moving outflows are tracked to the equatorial-layer outer radius $r_{\rm eq}$. The equatorial layer is represented by a finite-thickness gray absorbing layer, with off-equatorial plasma on both sides of the outflow-heated equatorial regions.

To describe an off-equatorial magnetic field concentrated near the equatorial layer and smoothly vanishing at its outer edge, we adopt a smooth combination of toroidal and poloidal distributions. The poloidal distribution $\widetilde{\mathcal B}_{\rm p}^i$ is constructed from $\Psi$ through Equation~\eqref{eq:flux}, while the toroidal distribution $\widetilde{\mathcal B}_{\rm t}^i$ has only an azimuthal flux component:
\begin{equation}
\begin{aligned}
\Psi&=-r_*^2\left(\frac r{r_*}\right)^{1/2}\sin^2\theta\,f_B(\theta)W_B(r),\\
\widetilde{\mathcal B}_{\rm t}^{\phi}
&=\frac1{r_*}\left(\frac r{r_*}\right)^{-5/2}f_B(\theta)W_B(r).
\end{aligned}
\label{eq:fieldshape}
\end{equation}

The angular envelope $f_B$ and radial window $W_B$ are
\begin{equation}
\begin{aligned}
f_B(\theta)&=\exp\!\left[-\frac{\cot^2\theta}{4h_B^2}\right],\\
W_B(r)&=\begin{cases}
\displaystyle\left(\frac{r_B-r}{r_B-r_*}\right)^2
\exp\!\left[\frac{2(r-r_*)}{r_B-r_*}\right],&r<r_B,\\
0,&r\geq r_B.
\end{cases}
\end{aligned}
\label{eq:fieldwindows}
\end{equation}
Here $h_B$ controls the angular concentration of the field, and $r_B$ is its outer radius.

The angular envelope is motivated by field-strength profiles obtained from Gaussian vertical density distributions and magnetic pressure scaling in radiatively inefficient accretion flow (RIAF) imaging models \cite{Broderick2011RIAF}. Since $\cot\theta=z/R$, it decreases with increasing height relative to cylindrical radius, while $\sin^2\theta$ makes the flux regular on the rotation axis. In the nonrelativistic weak-field limit, neglecting variations in the window functions, the two radial power laws in Equation~\eqref{eq:fieldshape} give physical poloidal and toroidal field strengths proportional to $r^{-3/2}$. Their magnetic pressure then has the same radial scaling as the outer background gas pressure. The radial window satisfies $W_B(r_*)=1$ and $W_B'(r_*)=0$, preserving the field at the reference point. At the outer edge, $W_B(r_B)=W_B'(r_B)=0$ ensures that the field vanishes continuously. Applying the window to the flux function before differentiation preserves the divergence-free poloidal field.

The reference point lies at radius $r_*$ and proper polar height $H_*$ above the equatorial plane. Let $N_{\rm p},N_{\rm t}$ be the comoving field norms of the two distributions at this point. After separate normalization, they are combined with the mixing angle $\zeta$ to give the off-equatorial magnetic flux variables
\begin{equation}
\mathcal B^i=B_*
\left(\frac{\sin\zeta}{N_{\rm p}}\widetilde{\mathcal B}_{\rm p}^i
+\frac{\cos\zeta}{N_{\rm t}}\widetilde{\mathcal B}_{\rm t}^i\right).
\label{eq:fieldmix}
\end{equation}
Here $B_*$ is a reference field amplitude satisfying $B_*^2=8\pi p_*/\beta_*$. The parameter $\beta_*$ sets the common normalization scale through $B_*$, while $\zeta$ controls the mixing ratio of the two distributions. With radial inflow, their comoving magnetic fields need not be orthogonal. The actual total field strength is calculated from Equation~\eqref{eq:bfield}, and the local plasma $\beta$ is determined from the combined field.

We adopt the mass, distance, and quiescent flux scales of M87$^*$ \cite{R17,EHTM87VI} and use the electron-temperature relation in Equation~\eqref{eq:temperature} to examine changes caused by heating. The same mass accretion rate $\dot M$ sets both the background radial inflow and the reconnection outflow power. The adopted parameters are listed in Table~\ref{tab:parameters}.

We adopt $\epsilon_b=1$, corresponding to $L_0\simeq5.66\times10^{43}\,\mathrm{erg\,s^{-1}}$, and use $\eta=f_{\rm off}=0.9$ to examine efficient thermalization and energy transfer to the off-equatorial regions. Of the energy already transferred from the outflow to the equatorial layer, $81\%$ enters the two off-equatorial regions as heat through Eq.~\eqref{eq:heating}. With $\ell=10r_g$ and $m=2$, the outflow has transferred $75\%$ of its initial energy to the equatorial layer after traveling a path length of $10r_g$. We take the dimensionless cooling parameter $\tau_c=2$, corresponding to a physical proper cooling time of $2t_g$. Heat retention in the off-equatorial plasma is also affected by the explicitly included advection and adiabatic evolution.

\begin{widefigure}
\centering
\includegraphics[width=\linewidth]{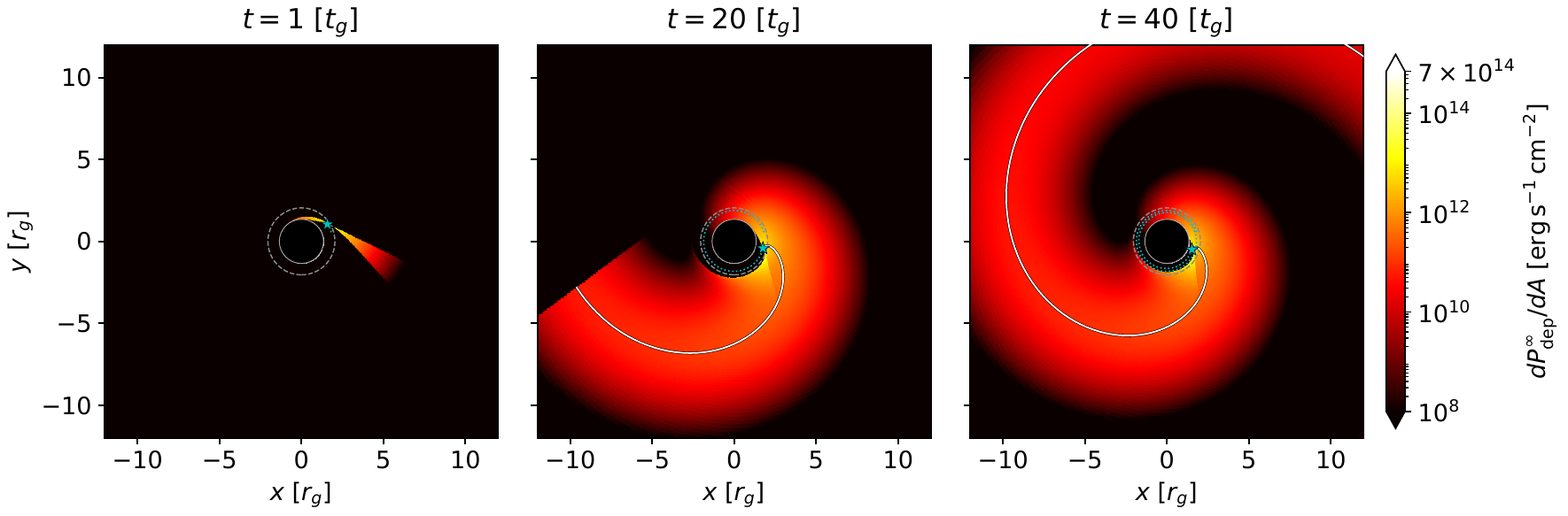}
\caption{Outflow central curves and deposited energy flux received by the equatorial layer at BL coordinate times $t=1,20,40t_g$, from left to right. Colors show $dP_{\rm dep}^{\infty}/dA$, where $dA=r_g^2\sqrt{g_{rr}g_{\phi\phi}}\,dr\,d\phi$ is the equatorial proper area element on a constant-BL-time slice. White solid curves trace the instantaneous outflow centers, cyan dotted curves show the past source trajectory, and stars mark the current source position. The black circle and gray dashed circle indicate the event horizon and ISCO, respectively. The display coordinates are $x=r\cos\phi,y=r\sin\phi$.}
\label{fig:outflow}
\end{widefigure}

\subsection{Spatial Evolution of the Outflow and Heated Source}
\label{sec:source}

Figure~\ref{fig:outflow} shows the outflow central curves and the deposited energy flux supplied to the equatorial layer. Continuous launching from the plunging source produces winding curves, and the finite transverse width spreads the energy input into bands along them. The loss function gives a larger energy transfer per unit arc length near the source, while outward-propagating outflows carry their remaining energy to larger disk radii. Strong energy deposition near the black hole therefore coexists with weaker extended bands, and the distribution changes azimuth as the source moves.

\begin{widefigure}
\centering
\includegraphics[width=\linewidth]{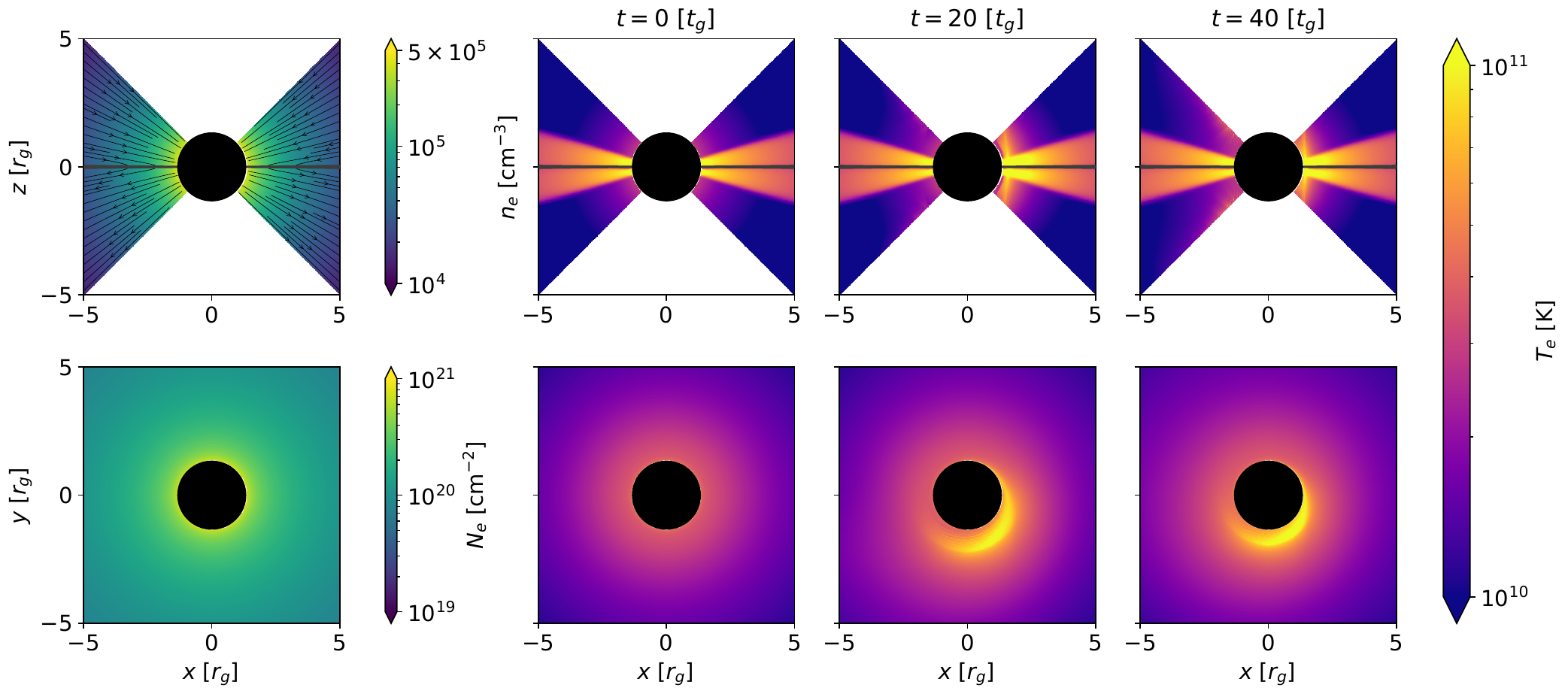}
\caption{Background density and electron temperature evolution of the thick-disk model. The top row shows the $xz$ plane at $y=0$; the bottom row projects the southern off-equatorial region onto the $xy$ plane by integrating along paths of fixed $R,\phi$. The first column gives the electron number density $n_e$ and column density $N_e=\int n_e\,d\ell$, where $d\ell$ is the proper line element along these paths on a constant-BL-time slice. Black curves with arrows in the upper-left panel trace the poloidal magnetic field. The remaining columns show the local temperature and number-density-weighted temperature $\int n_eT_e\,d\ell/N_e$ at source times $t=0,20,40t_g$. The display coordinates are $x=r\sin\theta\cos\phi$, $y=r\sin\theta\sin\phi$, and $z=r\cos\theta$. The black circle indicates the horizon, the narrow dark gray band is the equatorial layer, and white areas lie outside the emitting medium.}
\label{fig:heating}
\end{widefigure}

Figure~\ref{fig:heating} shows the background density and electron temperature at representative times. The density follows the same power law in Equation~\eqref{eq:background} throughout the background outside the horizon, with $q=2$. It therefore increases inward along a fixed polar angle and continues across the ISCO. Mass conservation determines the radial inflow velocity from the prescribed density, which remains fixed throughout the heating event.

Heating is concentrated within the central few $r_g$ and produces nonaxisymmetric hot arcs. In the column-averaged temperature of the southern off-equatorial region, the stronger heating at $t=20t_g$ and $40t_g$ lies near the black hole. The hot arc changes azimuth with time, while changes farther out are much weaker. This evolution reflects the thermal response excited at different positions by the moving energy input, together with advection of excess internal energy by the plasma. Propagation delays along the channels cause different heights to receive equatorial-layer energy at different times, while cooling continuously removes excess internal energy. Together, these processes determine the extent and duration of the heated structures.

The power transferred to the equatorial layer and the excess off-equatorial internal energy peak at $40.75t_g$ and $50.5t_g$, respectively. Energy received by the equatorial layer takes a finite time to reach the off-equatorial regions and is subsequently modified by advection, adiabatic changes, and cooling. The stored off-equatorial heat therefore depends on the previous heating history. This delays its maximum relative to the equatorial-layer power peak and gives the temperature evolution a different time dependence from the instantaneous outflow power.

\subsection{Polarized Imaging and Observables}
\label{sec:transfer}

We perform polarized radiative-transfer calculations separately for the quiescent and heated states using CoportSL \cite{CoportSL}. Rays are traced backward from the observer along null geodesics, with a parallel-transported polarization basis, and integration of the Stokes transfer equation yields the $I,Q,U,V$ images. The method is described in \cite{Huang2024Coport,R09}. We adopt isotropic thermal electrons with a Maxwell--J\"uttner distribution and the fitting formulae for thermal synchrotron emission, absorption, and Faraday coefficients in \cite{R11,R12}, using the high-temperature approximations for the Bessel functions. The Faraday rotation and conversion coefficients are denoted by $\rho_V,\rho_Q$, respectively.

Fast light fixes the source time throughout each image, while slow light samples the source at different times along each ray according to the propagation delay \cite{CoportSL}. We use the shortest delay window containing $99\%$ of the discrete transfer samples; samples outside it use the nearest endpoint time, and all samples contribute to the transfer integral. We determine the overall time shift $\Delta t_{\text{align}}$ from the cross-correlation of the fast- and slow-light total fluxes, and define the calibrated slow-light time as $t_{\rm obs}=t_{\rm frame}-\Delta t_{\text{align}}$, where $t_{\rm frame}$ labels the slow-light image. For this configuration, $\Delta t_{\text{align}}=27.25t_g$. Fast light uses the source time $t$; the same shift is applied to polarization, and comparisons use the common time interval. The observer direction and frequency are given in Table~\ref{tab:parameters}, and the imaging and timing settings in Table~\ref{tab:imaging}.

\begin{table}[!tbp]
\caption{Imaging and timing parameters.}
\label{tab:imaging}
\centering
\small
\setlength\tabcolsep{3pt}
\begin{tabular}{p{.21\linewidth}p{.32\linewidth}p{.37\linewidth}}
\toprule
Category & Setting & Value\\
\midrule
Camera & Camera radius & $500r_g$\\
& Field-of-view parameter (rad) & $\pi/64$\\
& Time-reference sphere radius & $20r_g$\\
Sampling & Image pixels & $512\times512$\\
& Image time interval & $0.25t_g$\\
Slow light & Transfer-sample coverage & $99\%$\\
\bottomrule
\end{tabular}
\end{table}

\begin{samepage}
We characterize the images by the total flux density, the image-averaged linear polarization (LP) fraction, and the electric vector position angle (EVPA) \cite{EHTM87VII,R09}:
\begin{equation}
\begin{aligned}
F_\nu&=\int I_\nu\,d\Omega,\\
m_{\rm L}&=\frac{\sum_p\sqrt{Q_p^2+U_p^2}}{\sum_p I_p},\\
\chi&=\frac12\operatorname{atan2}(U,Q)\pmod\pi.
\end{aligned}
\label{eq:ml}
\end{equation}
Here $p$ labels pixels of equal solid angle. The statistic $m_{\rm L}$ first forms the linear polarization intensity in each pixel and then sums these intensities, giving the intensity-weighted image average of the pixelwise LP fraction; $\chi$ is the EVPA at each pixel.
\end{samepage}

\section{Imaging Results}
\label{sec:results}

\subsection{Light Curves and Image Evolution}
\label{sec:lightcurve}

To establish the observable signal of localized reconnection heating, we compare light curves and images of the same evolving thermal source in fast and slow light. Figure~\ref{fig:lightcurve} shows the total flux density and image-averaged linear polarization fraction at $230\,\mathrm{GHz}$. Its horizontal axis is the source time $t$ for fast light and the calibrated time $t_{\rm obs}=t_{\rm frame}-\Delta t_{\text{align}}$ for slow light. The quiescent flux is approximately $0.658\,\mathrm{Jy}$, and the maximum fast- and slow-light fluxes are approximately $0.785\,\mathrm{Jy}$ and $0.822\,\mathrm{Jy}$, reaching $1.19$ and $1.25$ times the quiescent value, respectively. Both calculations use the same background and evolving thermal source. The treatment of light travel time changes the observed peak heights, shapes, and extrema times.

\begin{inlinefigure}
	\centering
	\includegraphics[width=\linewidth]{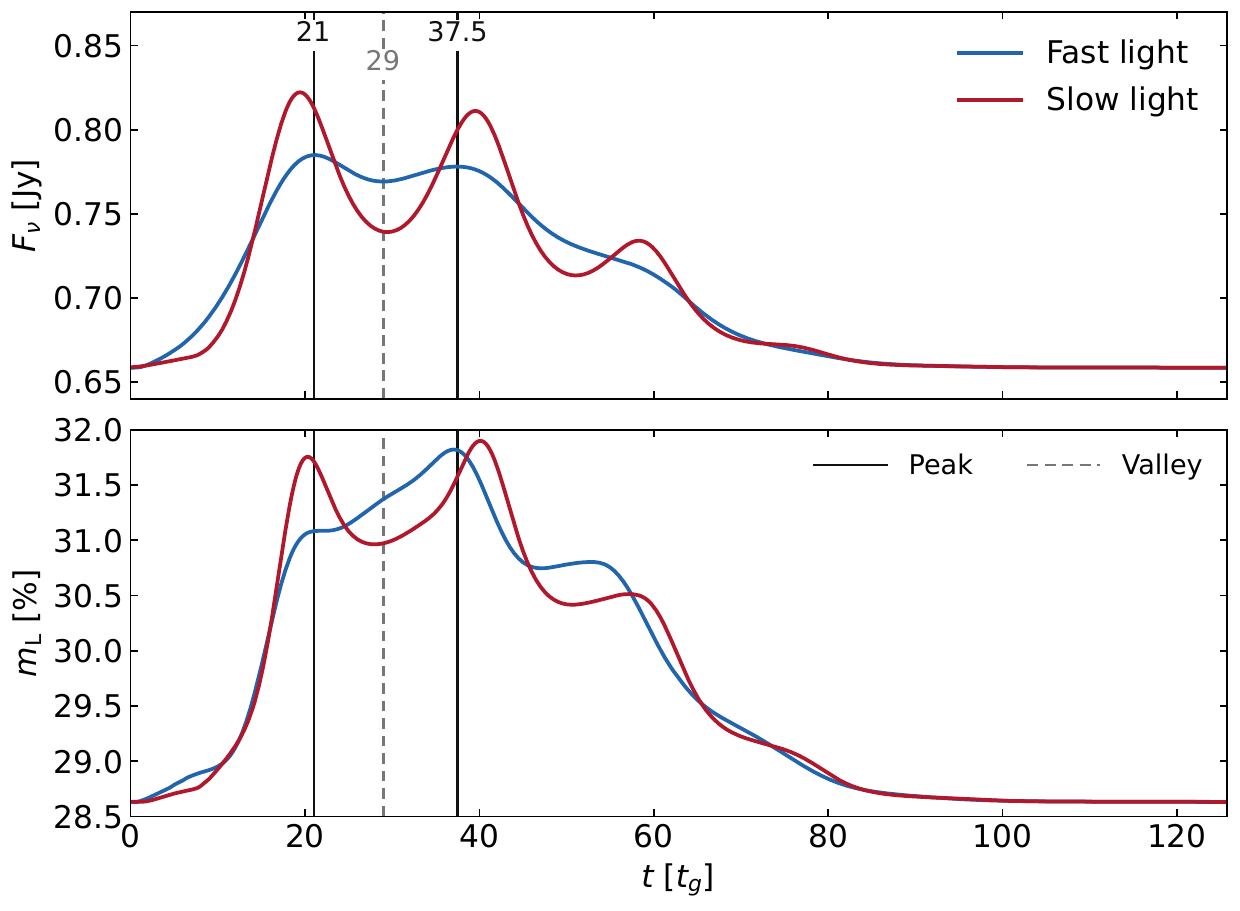}
	\caption{Time evolution of total flux density and image-averaged LP fraction. The top panel shows $F_\nu$ at $230\,\mathrm{GHz}$, and the bottom panel shows $m_{\rm L}$ defined in Equation~\eqref{eq:ml}. Blue and red curves denote fast and slow light, respectively. The slow-light time is aligned with fast light using the flux correlation. Black solid and gray dashed vertical lines mark the two fast-light total-flux peaks and one trough, respectively.}
	\label{fig:lightcurve}
\end{inlinefigure}

Fast light has two main flux peaks at $21t_g$ and $37.5t_g$, reaching approximately $0.785$ and $0.778\,\mathrm{Jy}$, respectively. Slow light has three peaks at $19.5t_g$, $39.5t_g$, and $58.25t_g$, reaching approximately $0.822$, $0.811$, and $0.734\,\mathrm{Jy}$, respectively. Fast light has one trough and slow light has two. The fast-light trough is approximately $0.769\,\mathrm{Jy}$ and remains above quiescence. The equatorial-layer power and excess off-equatorial internal energy peak at $40.75t_g$ and $50.5t_g$, respectively, as discussed in Section~\ref{sec:source}, and do not coincide directly with the light-curve extrema. Several observed extrema during sustained heating indicate that the mapping from energy input to variability is modulated by the spatial evolution of the heated structures and radiative propagation. Both the flux and image-averaged linear polarization fraction gradually return to quiescence as the flare decays.

To examine the emitting structures associated with the peaks and troughs, we compare representative images in Figure~\ref{fig:images}, using a sky position angle of $110^\circ$. The columns show the reference epoch $t=0$, the first fast-light flux peak at $21t_g$, the fast-light trough at $29t_g$, and the late flare stage at $70t_g$. Before heating, the image exhibits a central dark region surrounded by a ring with an uneven brightness distribution, morphologically similar to the millimeter image of M87$^*$~\cite{EHTM87I,EHTM87IV}. After heating, localized brightening is concentrated in arcs around the central dark region. Their brightness distributions and polarization directions evolve with the heated structures. Fast and slow light show similar bright-region locations at a given comparison time, while arc widths, details, and polarization directions change with the treatment of light travel time.

\begin{widefigure}
	\centering
	\includegraphics[width=\linewidth]{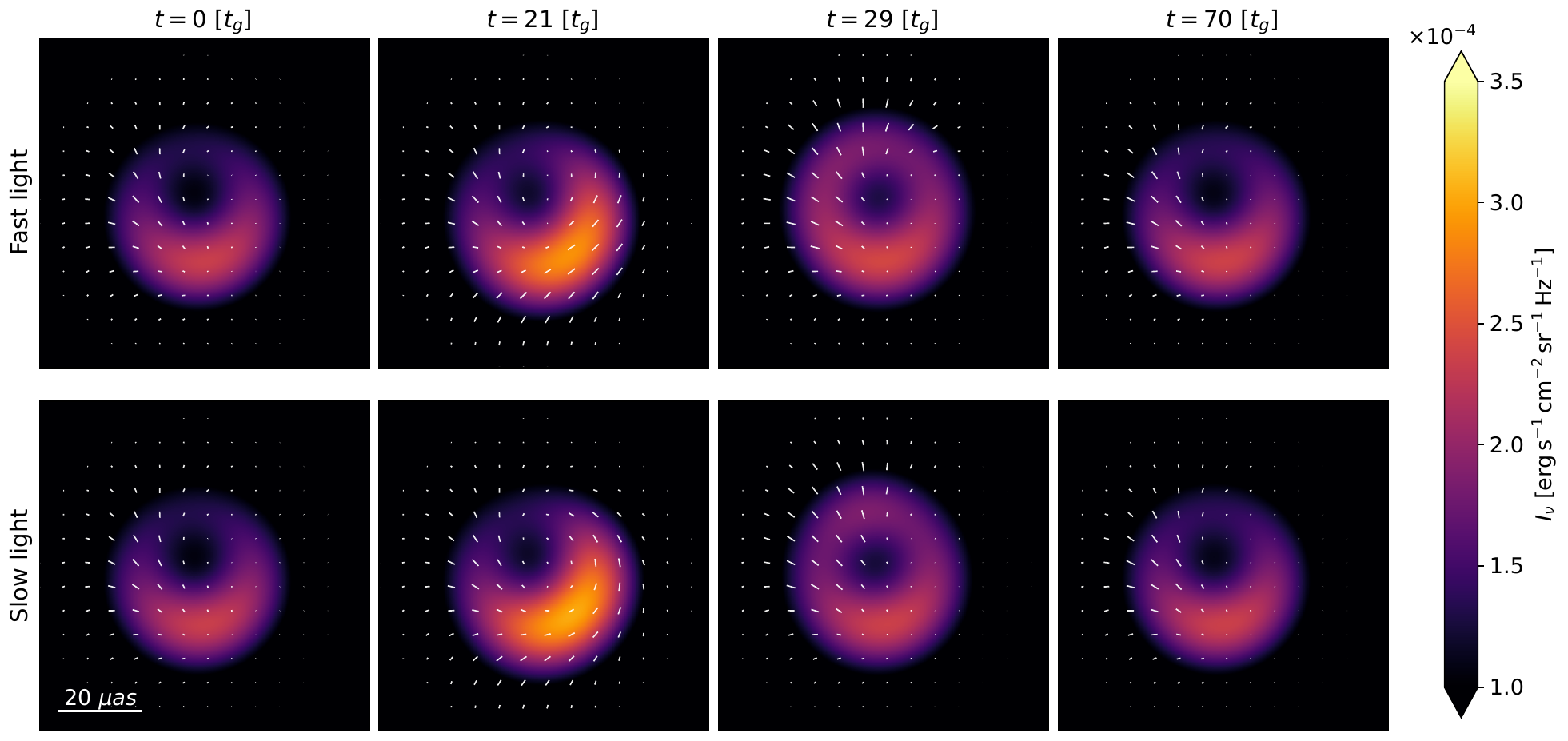}
	\caption{Intensity and EVPA at $230\,\mathrm{GHz}$. The four columns show the reference epoch $t=0$, the first fast-light flux peak, the first fast-light trough, and the late flare stage, respectively. The top and bottom rows show fast and slow light, respectively. Colors show the specific intensity $I_\nu$. White segment directions indicate EVPAs, and their lengths are normalized to the maximum quiescent linear polarization intensity. Each Stokes map is convolved with a Gaussian of FWHM $17\,\mu\mathrm{as}$.}
	\label{fig:images}
\end{widefigure}

\begin{widefigure}
	\centering
	\includegraphics[width=0.75\linewidth]{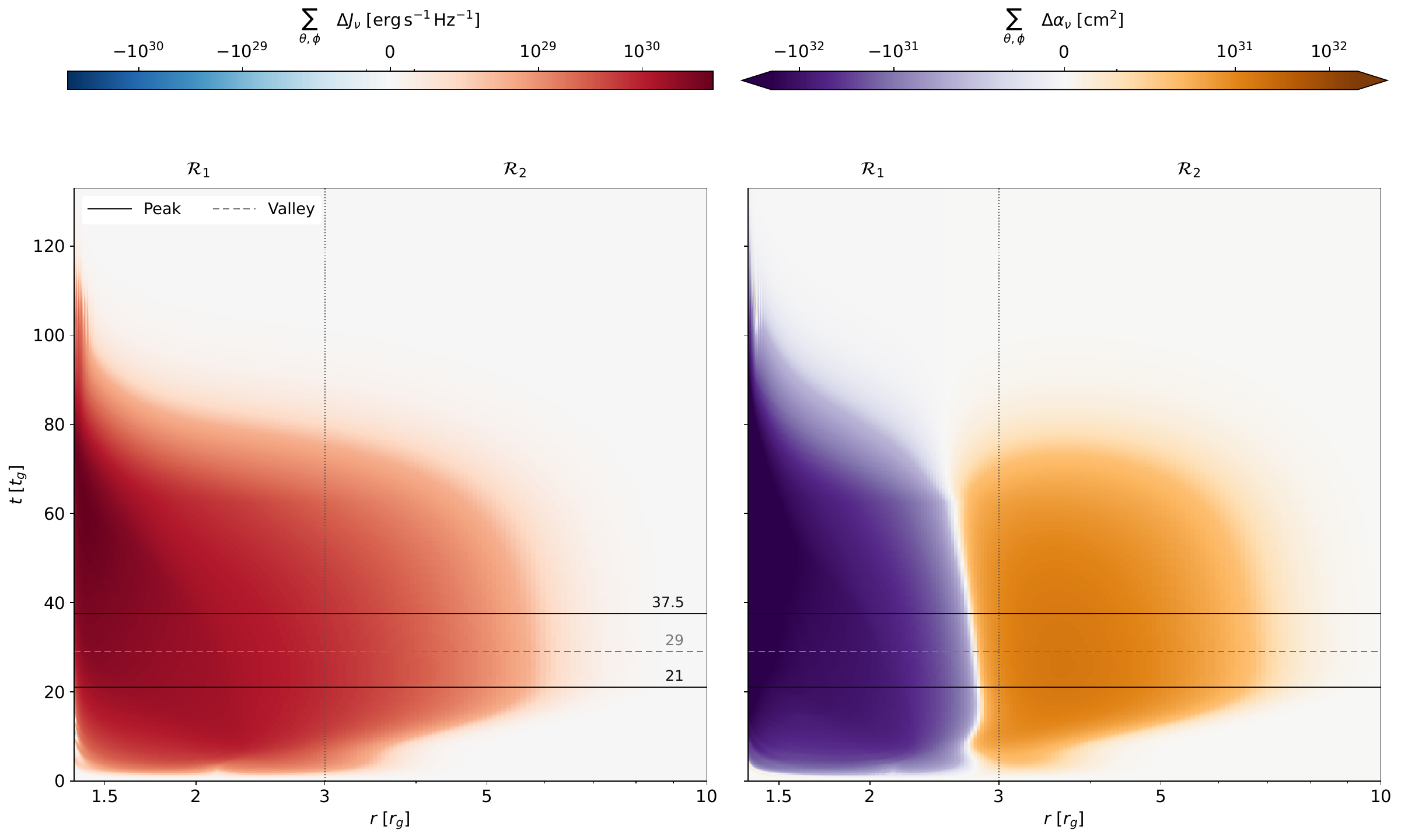}
	\caption{Radial differences in the $230\,\mathrm{GHz}$ emission and absorption coefficients from quiescence in the southern off-equatorial medium. The left and right panels show the direction-integrated emissivity difference $\Delta J_\nu=J_\nu-J_{\nu,0}$ and direction-averaged absorption coefficient difference $\Delta\alpha_\nu=\alpha_\nu-\alpha_{\nu,0}$, respectively, where subscript $0$ denotes quiescence. Both are evaluated in the comoving frame and integrated over $\theta,\phi$ with proper volume weights. The vertical dotted line at $r=3r_g$ separates the inner region $\mathcal R_1$ and outer region $\mathcal R_2$. Black solid and gray dashed horizontal lines mark the two fast-light flux peaks and one trough, respectively.}
	\label{fig:radial}
\end{widefigure}

\subsection{Radial Response and Regional Flux Contributions}
\label{sec:radial}

To locate the radial origins of the bright arcs and flux variations, we first examine the radial response of emission and absorption to heating. Let $J_\nu$ be the direction-integrated emissivity and $\alpha_\nu$ the direction-averaged absorption coefficient. We define the dimensional differences from quiescence as $\Delta J_\nu=J_\nu-J_{\nu,0}$ and $\Delta\alpha_\nu=\alpha_\nu-\alpha_{\nu,0}$, where subscript $0$ denotes quiescence. Figure~\ref{fig:radial} integrates these differences over polar and azimuthal angles with proper volume weights to display their radial evolution. Both coefficients are evaluated at a comoving frequency of $230\,\mathrm{GHz}$, with proper volume weights on constant-BL-time slices. The transfer calculation for the observed flux uses the actual comoving frequency and magnetic field angle along each ray.

We divide the southern off-equatorial medium into an inner region $\mathcal R_1:r\leq3r_g$ and an outer region $\mathcal R_2:3r_g<r\leq10r_g$, using fixed radial boundaries to compare their evolving responses in Figure~\ref{fig:radial}. During the main flare interval $10$--$60t_g$, the region-integrated emissivity increases in both regions, while integrated absorption decreases in the inner region and increases in the outer region. We refer to these as the emission-enhanced, absorption-reduced inner region and the region with enhanced emission and absorption, respectively. The absorption sign change moves between approximately $2.6$ and $2.8r_g$ with time, so this fixed partition describes the regional response; local changes still depend on azimuth and time. At fixed density and magnetic field, heating shifts the characteristic thermal synchrotron frequency upward, so the emissivity response at a fixed frequency depends on its position in the local spectrum. Thermal absorption satisfies $\alpha_I=j_I/B_\nu(T_e)$, with $B_\nu(T_e)\propto T_e$ in the Rayleigh--Jeans limit relevant here. Emission can increase while absorption decreases when the fractional emissivity increase is smaller than the temperature increase. A more rapid emissivity increase in the outer region can enhance both coefficients.

Emission and absorption can increase in the same region, so its observed contribution must be determined through the full radiative transfer. To connect the local responses to the observed flux, we retain the emission terms of one region at a time while including absorption, Faraday rotation, and conversion throughout the full medium. We denote the resulting flux by $C_i(t)$. Each region's emission thus experiences the same propagation conditions as in the full calculation, and emission contributions add linearly under the same transfer matrix. The contributions of the two regions and the remaining emission origins sum to the total flux, as shown in Figure~\ref{fig:masks}.

\begin{inlinefigure}
	\centering
	\includegraphics[width=0.76\linewidth]{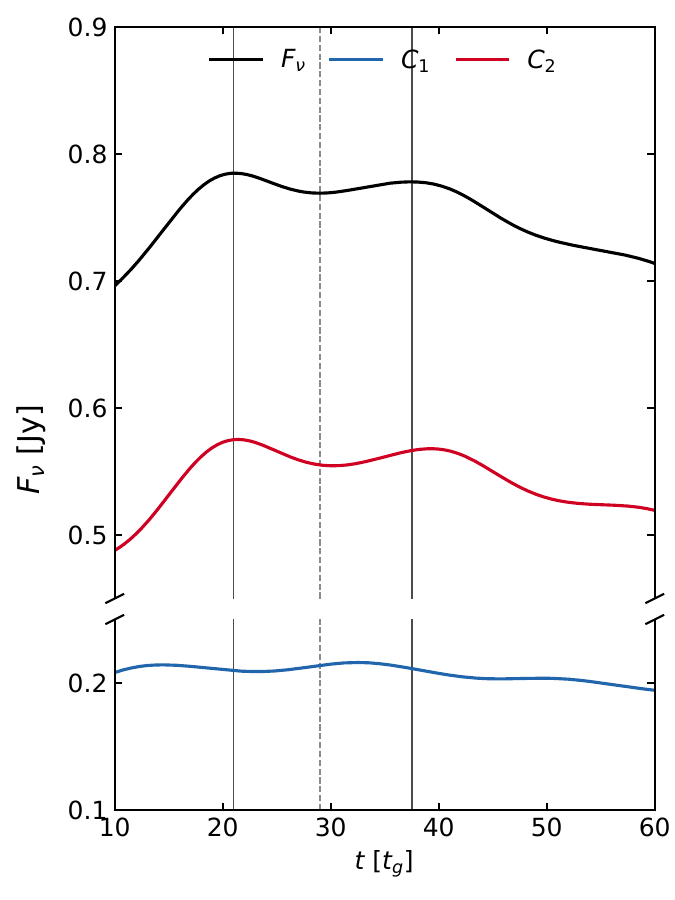}
	\caption{Fast-light flux contributions from two radial regions of the southern off-equatorial medium. Blue and red curves correspond to the inner region $\mathcal R_1:r\leq3r_g$ and outer region $\mathcal R_2:3r_g<r\leq10r_g$, respectively; the black curve gives the total flux. Each regional contribution retains emission only from that region, while absorption and Faraday transfer are retained throughout the full medium. Solid black and dashed gray vertical lines mark the two fast-light total-flux peaks and one trough, respectively.}
	\label{fig:masks}
\end{inlinefigure}

Throughout $10$--$60t_g$, the outer region $\mathcal R_2$ supplies the largest absolute flux, accounting for approximately $70\%$--$74\%$ of the total in fast and slow light. At the two fast-light flux peaks, $C_2$ is approximately $0.575$ and $0.567\,\mathrm{Jy}$, respectively, and approximately $0.555\,\mathrm{Jy}$ at the trough. The inner contributions at the same two peaks and trough are approximately $0.210$, $0.211$, and $0.214\,\mathrm{Jy}$. The peak-to-trough differences are therefore mainly expressed in the observed outer-region contribution. The reconnection source lies inside the ISCO, whereas the main observed emission comes from the more distant off-equatorial regions. Energy transport within the disk and radiative transfer along the rays are thus essential to connecting localized energy input to flare images. This decomposition also identifies the main target for the azimuthal analysis.

\subsection{Azimuthal Evolution and Foreground Absorption}
\label{sec:azimuth}

The radial decomposition locates the main observed emission, while interpreting the sequence of peaks and troughs also requires following the azimuthal evolution of the heated structures within the outer region $\mathcal R_2$.

Figure~\ref{fig:azimuth} integrates the emission and absorption changes over $r,\theta$, retaining their $\phi,t$ dependence. Enhanced emission and absorption form continuous slanted bands that connect across the $0^\circ/360^\circ$ boundary. At the two fast-light flux peaks, the emission enhancements reach azimuthal maxima near $302.4^\circ$ and $223.2^\circ$, respectively. The maximum at the trough is near $77.4^\circ$, showing that peaks and troughs correspond to different emitting azimuths.

\begin{widefigure}
	\centering
	\includegraphics[width=0.66\linewidth]{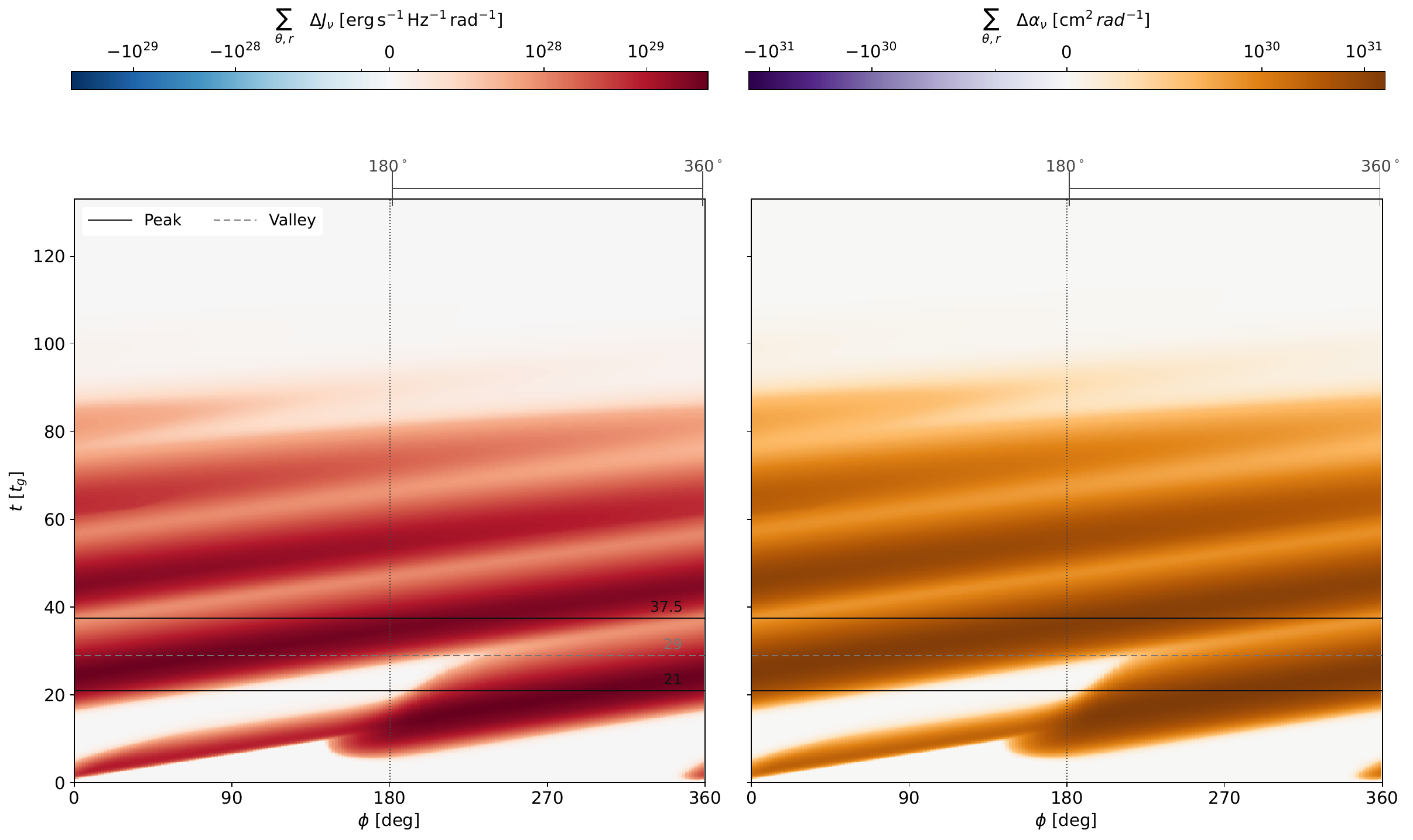}
	\caption{Azimuthal evolution of the emission difference $\Delta J_\nu$ and absorption difference $\Delta\alpha_\nu$ from quiescence within $\mathcal R_2$ ($3$--$10r_g$). Difference definitions follow Figure~\ref{fig:radial}; integration over $r,\theta$ retains $\phi,t$. Black solid and gray dashed horizontal lines mark the two fast-light total-flux peaks and one trough, respectively. Vertical dotted lines and the upper brackets indicate the azimuthal reference window $180^\circ$--$360^\circ$.}
	\label{fig:azimuth}
\end{widefigure}

To compare the emitting azimuths associated with peaks and troughs, we select $180^\circ$--$360^\circ$ as the azimuthal reference window. Keeping its width at $180^\circ$, we move its boundaries in $10^\circ$ steps to maximize the enclosed positive emission enhancement averaged over the two peak times. The window contains approximately $89\%$ and $72\%$ of the positive enhancement at those peaks and approximately $13\%$ at the trough. Its smaller enclosed fraction at the trough shows that the strongly emitting structures occupy different azimuths at peaks and troughs. This emission-selected window also provides a common azimuthal range for comparing foreground propagation conditions.

Because emission and absorption increase together in $\mathcal R_2$, we also examine absorption in front of the strongly emitting locations. At each time, we select the $10^\circ$ bin containing the strongest emission enhancement in that region. For each pixel ray crossing that bin, we choose the quadrature point with the largest unattenuated emission coefficient inside it as a representative source point. Integrating the absorption coefficient over the comoving path from that point to the observer gives the foreground optical depth $\tau_{\rm fore}$. The depth of the quiescent medium at the same source point provides a reference. Figure~\ref{fig:foreground} gives equal-pixel and equal-time averages within each azimuthal bin over $10$--$60t_g$.

\begin{inlinefigure}
	\centering
	\includegraphics[width=\linewidth]{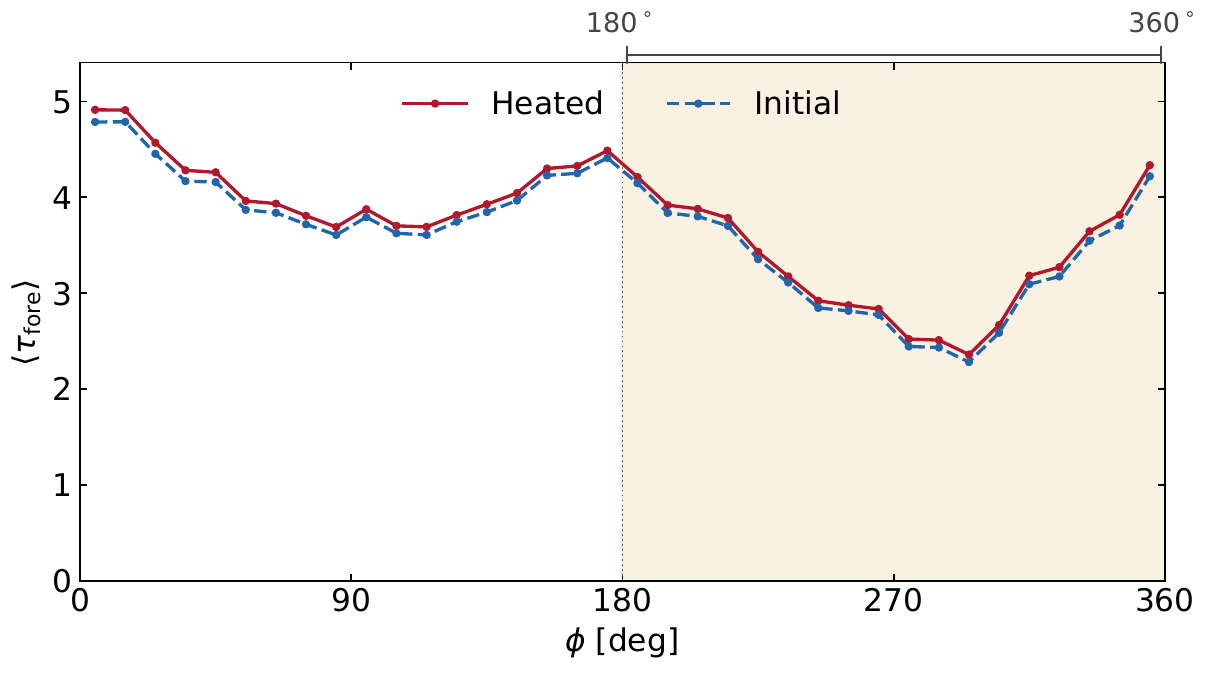}
	\caption{Mean foreground optical depth to the observer from strongly emitting regions in $\mathcal R_2$ ($3$--$10r_g$) versus azimuth, at fast-light times $10$--$60t_g$. At each time we select the $10^\circ$ bin of strongest emission enhancement and, on each ray, its point of highest unattenuated emissivity. Pixel--time samples are equally weighted within each bin. Red solid and blue dashed curves show the heated and quiescent media at the same source points. Shading and vertical dotted lines mark the $180^\circ$--$360^\circ$ reference window.}
	\label{fig:foreground}
\end{inlinefigure}

Grouping the above samples by whether the source azimuth lies inside or outside the reference window gives mean foreground optical depths of approximately $3.34$ and $4.10$, respectively, about $19\%$ lower inside the window. The corresponding medians are approximately $4.3\times10^{-3}$ and $3.8\times10^{-3}$. Their ordering differs from that of the means, reflecting the long tails of the depth distributions. The heated and quiescent-medium curves at the same source points have similar azimuthal variations, showing that the spatial structure of the background and the ray geometry already produce azimuth-dependent foreground absorption. The strongly emitting structures at the two fast-light flux peaks lie inside the window, where the mean foreground optical depth is lower and conditions are more favorable for radiation to escape; the strongly emitting structure at the trough lies mainly outside it. The azimuthal migration of the strongly emitting structures and the different foreground absorption they encounter provide a spatial picture for the flux peaks and troughs. Translating these propagation conditions into quantitative changes in peak flux also requires the source emissivity weights and transmission along the rays. Quiescent emission is approximately uniform in azimuth, and the azimuthal variation of the quiescent foreground optical depth shown in the figure indicates that differences in radiation escape are also a major cause of the brightness-asymmetric quiescent ring.

\subsection{Effects of Light Travel Time on Variability}
\label{sec:travel_time}

As the strongly emitting structures migrate through space, their light travel times to the observer change with position. To examine how neighboring emission intervals combine at the observer, we reuse the source points selected in Section~\ref{sec:azimuth} and read their geometric travel times. Let $\Delta t_\gamma$ be the travel time relative to the imaging reference travel time. On the comparison time axis of Section~\ref{sec:transfer},
\begin{equation}
t_{\rm obs}=t_{\rm em}+\Delta t_\gamma-\Delta t_{\text{align}} .
\label{eq:arrival}
\end{equation}

Here $t_{\rm em}$ is the emission time at the source point and $t_{\rm obs}$ is the calibrated slow-light time defined above. Figure~\ref{fig:travel_time} shows the mean arrival time and the $10$th--$90$th percentile interval of the source points at each emission time. A constant travel time gives a straight line with unit slope. Sections of the mean curve with slopes between $0$ and $1$ correspond to compression of emission-time intervals on the observer's time axis.

For example, the emission intervals $t_{\rm em}=15$--$20t_g$, $37$--$42t_g$, and $55$--$60t_g$, each lasting $5t_g$, map to mean arrival-time spans of approximately $1.98t_g$, $1.71t_g$, and $2.02t_g$, respectively. Neighboring emission states thus map to more closely spaced mean observer times during these intervals, allowing radiation to overlap more closely at the observer and enhancing slow-light peak-to-trough variability. Source points at any one emission time still have a $10$th--$90$th percentile arrival-time width of approximately $2.0$--$5.8t_g$. The finite arrival-time distributions and transfer along the rays jointly determine the observed peak times.

\begin{widefigure}
	\centering
	\includegraphics[width=0.8\linewidth]{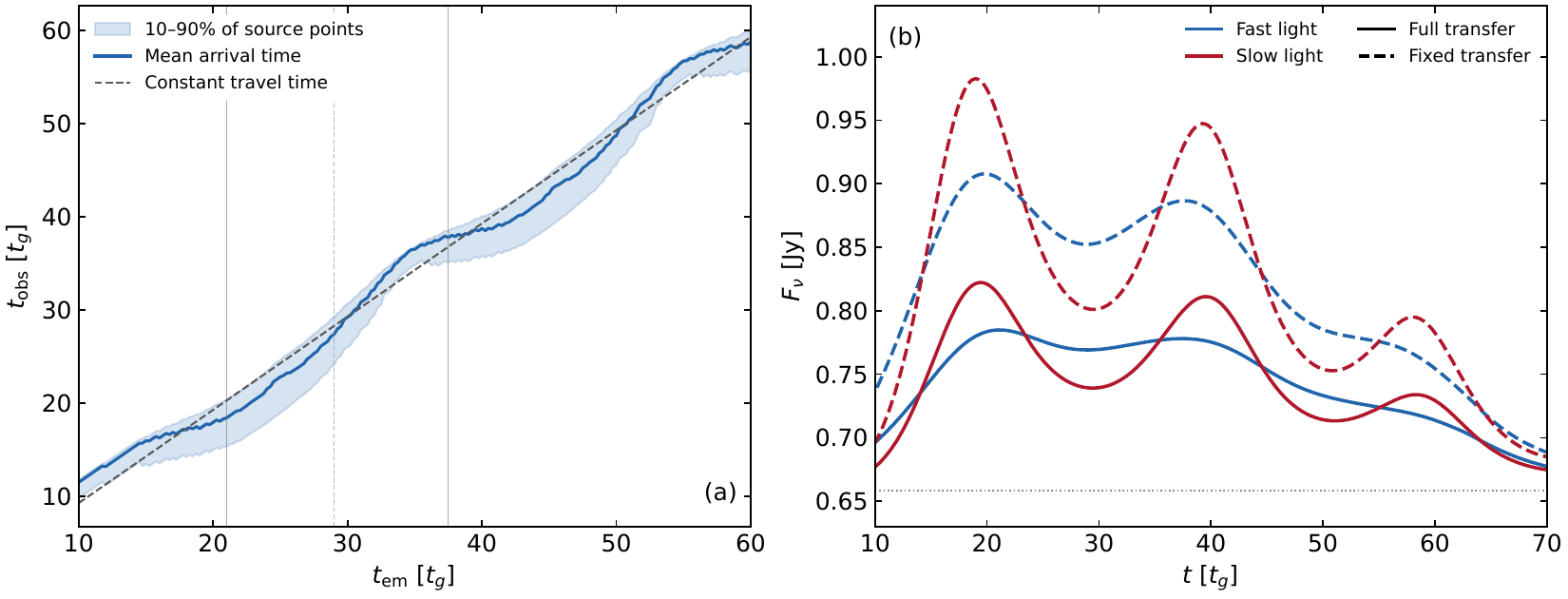}
	\caption{Left: emission and arrival times of strongly emitting source points in $\mathcal R_2$. At each time, we select the $10^\circ$ azimuthal bin with the strongest emission enhancement and, on each ray, take the point with the largest unattenuated emission coefficient within the bin as the representative source point. The blue curve and shading show the equally weighted mean arrival time and $10$th--$90$th percentile interval; the gray dashed line is a constant-travel-time reference. Solid and dashed vertical lines mark emission times of the fast-light flux peaks and trough. Right: total fluxes for the full calculation and the fixed-transfer control. Blue and red denote fast and slow light; solid and dashed curves denote full transfer and absorption and Faraday coefficients fixed at quiescence, respectively. The gray dotted line marks the quiescent flux. The horizontal axis is source time for fast light and calibrated time for slow light.}
	\label{fig:travel_time}
\end{widefigure}

To test the role of delayed sampling of evolving emission in the fast- and slow-light differences, we keep the thermal source, rays, and time alignment unchanged while updating only the four Stokes emissivity components and fixing absorption, Faraday rotation, and conversion coefficients at their quiescent values. The right panel of Figure~\ref{fig:travel_time} compares this control with the full calculation. The maximum fast- and slow-light fluxes in the control are approximately $0.908$ and $0.983\,\mathrm{Jy}$, or $1.38$ and $1.49$ times the quiescent value, respectively. Fast light still has two main peaks and slow light three. The flux drops from the first peak to the following trough are approximately $0.056$ and $0.182\,\mathrm{Jy}$, respectively, giving a larger peak-to-trough difference in slow light. Delayed sampling of evolving emission alone is therefore sufficient to produce the more pronounced slow-light variability. The lower maximum fluxes in the full calculation show that the thermal response of the propagation coefficients reduces the maximum brightening in this configuration.

\subsection{Linear Polarization Evolution and Line-of-Sight Cancellation}
\label{sec:polarization}

The image-averaged linear polarization fraction rises during the flare. In Figure~\ref{fig:lightcurve}, $m_{\rm L}$ increases from approximately $28.6\%$ in quiescence to maxima of $31.8\%$ in fast light and $31.9\%$ in slow light, at $37t_g$ and $40t_g$, respectively. These times are close to the corresponding second flux peaks at $37.5t_g$ and $39.5t_g$. The image-averaged linear polarization fraction gradually returns to quiescence as the flare decays. The flux and polarization maxima are not exactly simultaneous, reflecting different time evolution of emission brightening and cancellation among polarized contributions.

Heating changes both the weights of emission segments at different depths and polarized transfer along the rays. To examine cancellation between their observed contributions, we transport each emission segment to the observer separately within each pixel. Let $Q_{pk},U_{pk}$ be the observed Stokes contributions of emission segment $k$ in pixel $p$. We define the linear-polarization depolarization factor by
\begin{equation}
D_{\rm L}=1-
\frac{\displaystyle\sum_p\left|\sum_k(Q_{pk}+iU_{pk})\right|}
{\displaystyle\sum_{p,k}\sqrt{Q_{pk}^2+U_{pk}^2}} .
\label{eq:depolarization}
\end{equation}

\begin{inlinefigure}
\centering
\includegraphics[width=\linewidth]{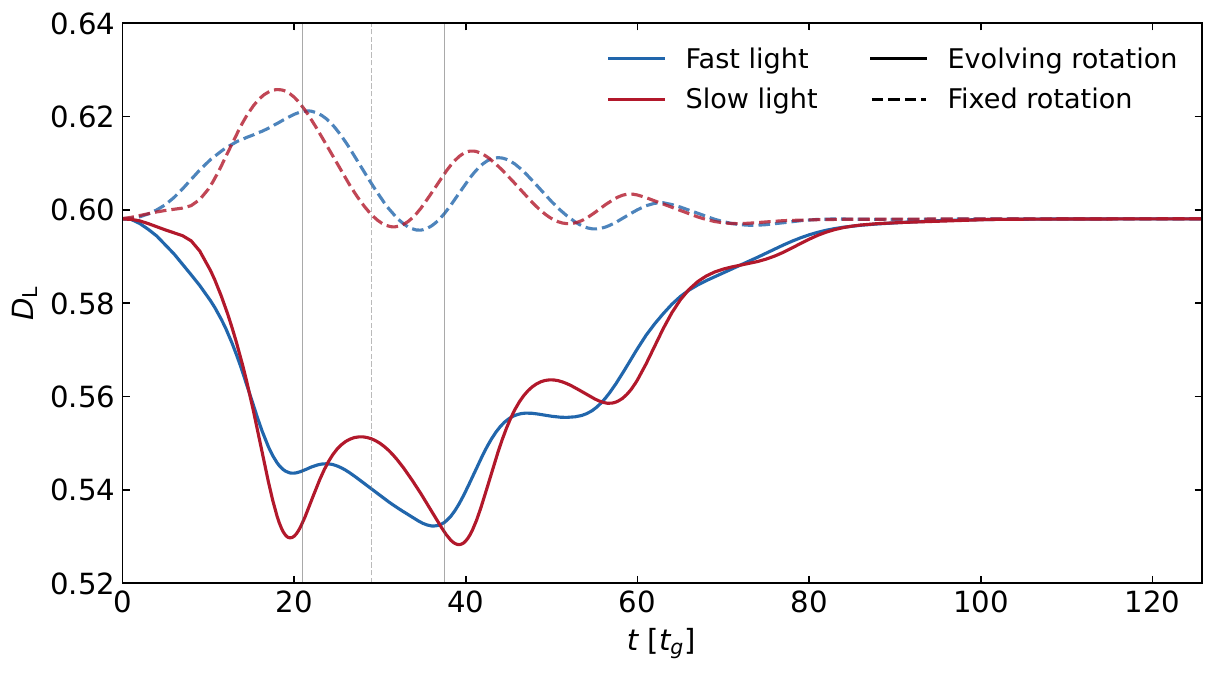}
\caption{Time evolution of the linear-polarization depolarization factor $D_{\rm L}$, defined in Equation~\eqref{eq:depolarization} to measure cancellation between polarization contributions from emission segments along the same pixel ray. Blue and red indicate fast and slow light. Solid and dashed curves show evolving rotation coefficients and coefficients fixed at their quiescent values, respectively. Vertical solid and dashed lines mark fast-light flux peaks and troughs, respectively.}
\label{fig:polarization_factors}
\end{inlinefigure}

Larger $D_{\rm L}$ corresponds to stronger cancellation between emission segments within the same pixel. To identify the thermal response of the rotation coefficient, Figure~\ref{fig:polarization_factors} compares full thermal evolution with a calculation holding $\rho_V$ at its quiescent value. Emission, absorption, and Faraday conversion continue to evolve with heating in the control calculation, retaining their effects on the addition of polarized contributions.

In the full calculation, $D_{\rm L}$ decreases from approximately $0.598$ in quiescence to minima of approximately $0.532$ in fast light and $0.528$ in slow light at $36.25t_g$ and $39.25t_g$, respectively. With fixed $\rho_V$, it varies between approximately $0.596$ and $0.626$, without a comparable decline. Because emission, absorption, and Faraday conversion continue to evolve with heating in the control, this result shows that the thermal response of Faraday rotation dominates the reduction of line-of-sight polarization cancellation during the flare in this configuration, promoting the increase in the image-averaged linear polarization fraction. Temperature changes modify the relative directions of contributions from different depths through the rotation coefficient, allowing more coherent addition along a given pixel ray.

\section{Conclusions and Discussion}
\label{sec:discussion}
\label{sec:conclusion}

Using a semianalytical thick-disk model, we have connected reconnection outflow power inside the ISCO to energy transport within the disk, thermal perturbations, and polarized radiative transfer. A Kerr configuration with a fixed background illustrates the millimeter flare produced by localized heating. The main results are as follows.

\begin{enumerate}
\item For the mass and distance of M87$^*$, the quiescent $230\,\mathrm{GHz}$ image shows a central brightness depression surrounded by a ring that is brighter on one side, as in the EHT image~\cite{EHTM87I,EHTM87IV}. When the reconnection outflow supplies energy to the equatorial layer and heat reaches the off-equatorial regions, arcs of the ring brighten markedly (Figure~\ref{fig:images}). This brightening occurs with the background density, velocity, and magnetic field held fixed, suggesting that localized magnetic reconnection could power transient brightening in the millimeter ring of M87$^*$. The peak total fluxes reach approximately $1.19$ and $1.25$ times the quiescent value in fast and slow light, respectively. One sustained heating episode produces two main peaks in fast light and three in slow light, showing that thermal evolution and radiative propagation can turn continuous energy input into an observed signal with several extrema.
\item The energy input and the main observed emission occupy different regions. In terms of region-integrated responses, the inner region $\mathcal R_1:r\leq3r_g$ shows predominantly enhanced emission and reduced absorption, while both increase in the outer region $\mathcal R_2:3r_g<r\leq10r_g$. The reconnection source lies inside the ISCO, whereas the outer region supplies approximately $70\%$--$74\%$ of the total flux. Strongly emitting structures migrate in azimuth within this region. At the two fast-light peaks, their emitting azimuths lie in the range of lower mean foreground optical depth, which favors radiation escape; the emitting azimuth at the trough lies mainly outside this range.
\item Motion of the strongly emitting structures changes light travel times with emission position. The mean arrival-time intervals of several emission periods are compressed, allowing neighboring emission states to overlap more closely at the observer and enhancing slow-light peak-to-trough variability. With absorption and Faraday coefficients fixed, slow light retains an additional main peak and a deeper trough relative to fast light, showing that delayed sampling of evolving emission is sufficient to produce this difference in the present configuration. The lower maximum fluxes in the full calculation show that the thermal response of the propagation coefficients also modifies the brightening amplitude.
\item The image-averaged linear polarization fraction rises from approximately $28.6\%$ to $31.8\%$--$31.9\%$ during the flare, with maxima close to the second flux peaks. The depolarization factor along pixel rays decreases during the same period. Holding the rotation coefficient at its quiescent value substantially weakens this trend, showing that the thermal response of Faraday rotation dominates the reduction of line-of-sight polarization cancellation in this configuration. Brightness and polarization thus reflect different aspects of the heated plasma's evolution and radiative transfer.
\end{enumerate}

The present configuration adopts one set of energy-input and thermal-response parameters. Varying the power parameter $\epsilon_b$ within the magnetic energy budget, together with the deposition length $\ell$ and cooling time $\tau_c$, would distinguish how energy input, spatial deposition, and heat retention affect flare amplitudes and durations. This would also establish the conditions under which successive peaks occur for different heating strengths and thermal-response times. Varying black hole spin, the initial source position, and viewing inclination would probe how the heated trajectories, peak separations, and arrival-time mapping depend on event geometry and viewing direction, testing the applicability of the azimuthal preference and enhanced slow-light variability found here.

Imaging at multiple frequencies would exploit the different frequency responses of emission, absorption, and Faraday effects to examine changes in the main emitting region, flux-peak times, and the timing of polarization enhancement. These comparisons would further distinguish thermal evolution at the source from foreground transfer. Combined variations of the magnetic mixing angle and electron temperature prescription would test the dependence of polarization enhancement on the ordered field geometry and the adopted electron-temperature relation. Including two-temperature and nonthermal electrons, together with synchrotron and Compton feedback on the energy evolution, would clarify how the partition of reconnection energy between thermal electron heating and nonthermal particle acceleration affects light curves and polarization \cite{R15,R19,R09}.

For numerical validation, the slow-light delay window can be extended to test the influence of the current finite window on peak times, peak amplitudes, and polarization evolution, providing numerical support for comparisons across physical conditions.

The model specifies the outflow power, upstream magnetization, and background imaging field separately, while fixing density, velocity, magnetic field, and disk geometry. Coupling reconnection power to disk dynamics, vertical equilibrium, and radiative cooling would allow the pressure response, density changes, and magnetic field evolution caused by heating to modify the net variability. This would connect the heating contribution isolated here to the global accretion-flow response during magnetic flux eruptions and establish the conditions under which pronounced brightening persists in the presence of dynamical backreaction.

\begin{acknowledgments}

We thank Xiaoxiong Zeng for helpful discussions. This work is partly supported by NSFC Grant Nos.\ 12575048, 12275004, and 12588101. M.G. also acknowledges support from the BNU Tang Scholar Program. B.C. was supported by the Ningbo Science and Technology
Bureau under Grant 2024DJRC010009.

\end{acknowledgments}

\bibliographystyle{apsrev4-2}
\bibliography{references}

@misc{Wan:2026aio,
    author = "Wan, Qian and Hou, Yehui and Song, Yu and Xia, Jing-Ze and Mizuno, Yosuke and Chen, Bin",
    title = "{An Electromagnetic Ergoregion Limits the Reconnection-Driven Penrose Process}",
    eprint = "2611.33995",
    archivePrefix = "arXiv",
    primaryClass = "gr-qc",
    month = "11",
    year = "2026"
}

@article{Shen:2024sdr,
    author = "Shen, Ye and YuChih, Ho-Yun and Chen, Bin",
    title = "{Energy extraction from a rotating black hole via magnetic reconnection: The plunging bulk plasma and orientation angle}",
    eprint = "2409.07345",
    archivePrefix = "arXiv",
    primaryClass = "gr-qc",
    doi = "10.1103/PhysRevD.110.123010",
    journal = "Phys. Rev. D",
    volume = "110",
    number = "12",
    pages = "123010",
    year = "2024"
}

@misc{R01,
  author = {Li, Zhen},
  title = {{Magnetic Reconnection Heating of Thin Accretion Disks around Kerr Black Holes}},
  year = {2026},
  eprint = {2609.15449},
  archivePrefix = {arXiv},
  note = {arXiv:2609.15449}
}

@article{R03,
  author = {James M. Bardeen and William H. Press and Saul A. Teukolsky},
  title = {{Rotating Black Holes: Locally Nonrotating Frames, Energy Extraction, and Scalar Synchrotron Radiation}},
  year = {1972},
  journal = {The Astrophysical Journal},
  volume = {178},
  pages = {347},
  doi = {10.1086/151796}
}

@article{R04,
  author = {Comisso, Luca and Asenjo, Felipe A.},
  title = {{Magnetic reconnection as a mechanism for energy extraction from rotating black holes}},
  year = {2021},
  journal = {Physical Review D},
  volume = {103},
  pages = {023014},
  doi = {10.1103/PhysRevD.103.023014}
}

@article{Asenjo2017Kerr,
  author = {Asenjo, Felipe A. and Comisso, Luca},
  title = {{Relativistic Magnetic Reconnection in Kerr Spacetime}},
  journal = {Physical Review Letters},
  volume = {118},
  pages = {055101},
  year = {2017},
  doi = {10.1103/PhysRevLett.118.055101}
}

@article{Comisso2018Collisionless,
  author = {Comisso, Luca and Asenjo, Felipe A.},
  title = {{Collisionless magnetic reconnection in curved spacetime and the effect of black hole rotation}},
  journal = {Physical Review D},
  volume = {97},
  pages = {043007},
  year = {2018},
  doi = {10.1103/PhysRevD.97.043007}
}

@article{Asenjo2019Gravity,
  author = {Asenjo, Felipe A. and Comisso, Luca},
  title = {{Gravitational electromotive force in magnetic reconnection around Schwarzschild black holes}},
  journal = {Physical Review D},
  volume = {99},
  pages = {063017},
  year = {2019},
  doi = {10.1103/PhysRevD.99.063017}
}

@article{R05,
  author = {Ripperda, Bart and Bacchini, Fabio and Philippov, Alexander},
  title = {{Magnetic Reconnection and Hot Spot Formation in Black Hole Accretion Disks}},
  year = {2020},
  journal = {The Astrophysical Journal},
  volume = {900},
  pages = {100},
  doi = {10.3847/1538-4357/ababab}
}

@article{R06,
  author = {Ripperda, Bart and Liska, Matthew and Chatterjee, Koushik and Musoke, Gibwa and Philippov, Alexander A. and Markoff, Sera B. and Tchekhovskoy, Alexander and Younsi, Ziri},
  title = {{Black Hole Flares: Ejection of Accreted Magnetic Flux through 3D Plasmoid-mediated Reconnection}},
  year = {2022},
  journal = {The Astrophysical Journal Letters},
  volume = {924},
  pages = {L32},
  doi = {10.3847/2041-8213/ac46a1}
}

@article{Bransgrove2021MagneticHair,
  author = {Bransgrove, Ashley and Ripperda, Bart and Philippov, Alexander},
  title = {{Magnetic Hair and Reconnection in Black Hole Magnetospheres}},
  year = {2021},
  journal = {Physical Review Letters},
  volume = {127},
  pages = {055101},
  doi = {10.1103/PhysRevLett.127.055101}
}

@article{Galishnikova2023Collisionless,
  author = {Galishnikova, Alisa and Philippov, Alexander and Quataert, Eliot and Bacchini, Fabio and Parfrey, Kyle and Ripperda, Bart},
  title = {{Collisionless Accretion onto Black Holes: Dynamics and Flares}},
  year = {2023},
  journal = {Physical Review Letters},
  volume = {130},
  pages = {115201},
  doi = {10.1103/PhysRevLett.130.115201}
}

@article{Jia2023MillimeterFlares,
  author = {Jia, He and Ripperda, Bart and Quataert, Eliot and White, Christopher J. and Chatterjee, Koushik and Philippov, Alexander and Liska, Matthew},
  title = {{Millimeter observational signatures of flares in magnetically arrested black hole accretion models}},
  year = {2023},
  journal = {Monthly Notices of the Royal Astronomical Society},
  volume = {526},
  pages = {2924--2941},
  doi = {10.1093/mnras/stad2935}
}

@misc{R07,
  author = {Zeng, Xiao-Xiong and Hong, Yun and Wang, Ke},
  title = {{Hotspot Images from Magnetic Reconnection Processes in the plunging Region of a Kerr Black Hole}},
  year = {2026},
  eprint = {2602.21836},
  archivePrefix = {arXiv},
  note = {arXiv:2602.21836}
}

@article{R08,
  author = {Zhao, Zhixing and Fan, Zhong-Ying and Wang, Xiaobao and Guo, Minyong and Chen, Bin},
  title = {{Probing the Penrose process: Images of split hotspots and their observational signatures}},
  year = {2026},
  journal = {Physical Review D},
  volume = {113},
  pages = {044019},
  doi = {10.1103/7mcs-7yb3}
}

@article{R09,
  author = {Zhou, Fan and Huang, Jiewei and Li, Yuehang and Zhang, Zhenyu and Hou, Yehui and Guo, Minyong and Chen, Bin},
  title = {{Nonthermal Synchrotron Emission and Polarization Signatures during Black Hole Flux Eruptions}},
  year = {2026},
  journal = {The Astrophysical Journal},
  volume = {1002},
  pages = {152},
  doi = {10.3847/1538-4357/ae5e67}
}

@article{R11,
  author = {Dexter, Jason},
  title = {{A public code for general relativistic, polarised radiative transfer around spinning black holes}},
  year = {2016},
  journal = {Monthly Notices of the Royal Astronomical Society},
  volume = {462},
  pages = {115--136},
  doi = {10.1093/mnras/stw1526}
}

@article{R12,
  author = {Pandya, Alex and Zhang, Zhaowei and Chandra, Mani and Gammie, Charles F.},
  title = {{Polarized synchrotron emissivities and absorptivities for relativistic thermal, power-law, and kappa distribution functions}},
  year = {2016},
  journal = {The Astrophysical Journal},
  volume = {822},
  pages = {34},
  doi = {10.3847/0004-637X/822/1/34}
}

@article{R15,
  author = {Chael, Andrew and Rowan, Michael E. and Narayan, Ramesh and Johnson, Michael D. and Sironi, Lorenzo},
  title = {{The role of electron heating physics in images and variability of the Galactic Centre black hole Sagittarius A*}},
  year = {2018},
  journal = {Monthly Notices of the Royal Astronomical Society},
  volume = {478},
  pages = {5209--5229},
  doi = {10.1093/mnras/sty1261}
}

@article{R17,
  author = {Akiyama, Kazunori and Alberdi, Antxon and Alef, Walter and others},
  title = {{First M87 Event Horizon Telescope Results. V. Physical Origin of the Asymmetric Ring}},
  year = {2019},
  journal = {The Astrophysical Journal Letters},
  volume = {875},
  pages = {L5},
  doi = {10.3847/2041-8213/ab0f43}
}

@article{R19,
  author = {Rowan, Michael E. and Sironi, Lorenzo and Narayan, Ramesh},
  title = {{Electron and Proton Heating in Transrelativistic Magnetic Reconnection}},
  year = {2017},
  journal = {The Astrophysical Journal},
  volume = {850},
  pages = {29},
  doi = {10.3847/1538-4357/aa9380}
}

@article{R22,
  author = {I. Raginski and Ari Laor},
  title = {{AGN coronal emission models -- I. The predicted radio emission}},
  year = {2016},
  journal = {Monthly Notices of the Royal Astronomical Society},
  volume = {459},
  pages = {2082--2096},
  doi = {10.1093/mnras/stw772}
}

@article{R02,
  author = {Don N. Page and Kip S. Thorne},
  title = {{Disk-Accretion onto a Black Hole. Time-Averaged Structure of Accretion Disk}},
  journal = {The Astrophysical Journal},
  volume = {191},
  pages = {499},
  year = {1974},
  doi = {10.1086/152990}
}

@article{R25,
  author = {Mummery, Andrew and Balbus, Steven},
  title = {{Accretion within the innermost stable circular orbit: analytical thermodynamic solutions in the adiabatic limit}},
  journal = {Mon.Not.Roy.Astron.Soc.},
  volume = {521},
  pages = {2439--2463},
  year = {2023},
  doi = {10.1093/mnras/stad641}
}

@article{R26,
  author = {Mummery, Andrew and Ingram, Adam and Davis, Shane and Fabian, Andrew},
  title = {{Continuum emission from within the plunging region of black hole discs}},
  journal = {Mon.Not.Roy.Astron.Soc.},
  volume = {531},
  pages = {366--386},
  year = {2024},
  doi = {10.1093/mnras/stae1160}
}

@misc{CoportSL,
  author = {Zhou, Fan and Huang, Jiewei and Li, Yuehang and Guo, Minyong and Chen, Bin},
  title = {{CoportSL: A Contribution-constrained Hybrid Slow-light Framework for Time-dependent Polarized GRMHD Imaging}},
  year = {2026},
  eprint = {2609.06607},
  archivePrefix = {arXiv},
  note = {arXiv:2609.06607}
}

@article{EHTM87I,
  author = {Akiyama, Kazunori and Alberdi, Antxon and Alef, Walter and others},
  title = {{First M87 Event Horizon Telescope Results. I. The Shadow of the Supermassive Black Hole}},
  journal = {The Astrophysical Journal Letters},
  volume = {875},
  pages = {L1},
  year = {2019},
  doi = {10.3847/2041-8213/ab0ec7}
}

@article{EHTM87II,
  author = {Akiyama, Kazunori and Alberdi, Antxon and Alef, Walter and others},
  title = {{First M87 Event Horizon Telescope Results. II. Array and Instrumentation}},
  journal = {The Astrophysical Journal Letters},
  volume = {875},
  pages = {L2},
  year = {2019},
  doi = {10.3847/2041-8213/ab0c96}
}

@article{EHTM87III,
  author = {Akiyama, Kazunori and Alberdi, Antxon and Alef, Walter and others},
  title = {{First M87 Event Horizon Telescope Results. III. Data Processing and Calibration}},
  journal = {The Astrophysical Journal Letters},
  volume = {875},
  pages = {L3},
  year = {2019},
  doi = {10.3847/2041-8213/ab0c57}
}

@article{EHTM87IV,
  author = {Akiyama, Kazunori and Alberdi, Antxon and Alef, Walter and others},
  title = {{First M87 Event Horizon Telescope Results. IV. Imaging the Central Supermassive Black Hole}},
  journal = {The Astrophysical Journal Letters},
  volume = {875},
  pages = {L4},
  year = {2019},
  doi = {10.3847/2041-8213/ab0e85}
}

@article{EHTM87VI,
  author = {Akiyama, Kazunori and Alberdi, Antxon and Alef, Walter and others},
  title = {{First M87 Event Horizon Telescope Results. VI. The Shadow and Mass of the Central Black Hole}},
  journal = {The Astrophysical Journal Letters},
  volume = {875},
  pages = {L6},
  year = {2019},
  doi = {10.3847/2041-8213/ab1141}
}

@article{EHTM87VII,
  author = {Akiyama, Kazunori and Algaba, Juan Carlos and Alberdi, Antxon and others},
  title = {{First M87 Event Horizon Telescope Results. VII. Polarization of the Ring}},
  journal = {The Astrophysical Journal Letters},
  volume = {910},
  pages = {L12},
  year = {2021},
  doi = {10.3847/2041-8213/abe71d}
}

@article{EHTM87VIII,
  author = {Akiyama, Kazunori and Algaba, Juan Carlos and Alberdi, Antxon and others},
  title = {{First M87 Event Horizon Telescope Results. VIII. Magnetic Field Structure near The Event Horizon}},
  journal = {The Astrophysical Journal Letters},
  volume = {910},
  pages = {L13},
  year = {2021},
  doi = {10.3847/2041-8213/abe4de}
}

@article{EHTM87IX,
  author = {Akiyama, Kazunori and Alberdi, Antxon and Alef, Walter and others},
  title = {{First M87 Event Horizon Telescope Results. IX. Detection of Near-horizon Circular Polarization}},
  journal = {The Astrophysical Journal Letters},
  volume = {957},
  pages = {L20},
  year = {2023},
  doi = {10.3847/2041-8213/acff70}
}

@article{EHTSgrAI,
  author = {Akiyama, Kazunori and Alberdi, Antxon and Alef, Walter and others},
  title = {{First Sagittarius A* Event Horizon Telescope Results. I. The Shadow of the Supermassive Black Hole in the Center of the Milky Way}},
  journal = {The Astrophysical Journal Letters},
  volume = {930},
  pages = {L12},
  year = {2022},
  doi = {10.3847/2041-8213/ac6674}
}

@article{EHTSgrAII,
  author = {Akiyama, Kazunori and Alberdi, Antxon and Alef, Walter and others},
  title = {{First Sagittarius A* Event Horizon Telescope Results. II. EHT and Multiwavelength Observations, Data Processing, and Calibration}},
  journal = {The Astrophysical Journal Letters},
  volume = {930},
  pages = {L13},
  year = {2022},
  doi = {10.3847/2041-8213/ac6675}
}

@article{EHTSgrAIII,
  author = {Akiyama, Kazunori and Alberdi, Antxon and Alef, Walter and others},
  title = {{First Sagittarius A* Event Horizon Telescope Results. III. Imaging of the Galactic Center Supermassive Black Hole}},
  journal = {The Astrophysical Journal Letters},
  volume = {930},
  pages = {L14},
  year = {2022},
  doi = {10.3847/2041-8213/ac6429}
}

@article{EHTSgrAIV,
  author = {Akiyama, Kazunori and Alberdi, Antxon and Alef, Walter and others},
  title = {{First Sagittarius A* Event Horizon Telescope Results. IV. Variability, Morphology, and Black Hole Mass}},
  journal = {The Astrophysical Journal Letters},
  volume = {930},
  pages = {L15},
  year = {2022},
  doi = {10.3847/2041-8213/ac6736}
}

@article{EHTSgrAV,
  author = {Akiyama, Kazunori and Alberdi, Antxon and Alef, Walter and others},
  title = {{First Sagittarius A* Event Horizon Telescope Results. V. Testing Astrophysical Models of the Galactic Center Black Hole}},
  journal = {The Astrophysical Journal Letters},
  volume = {930},
  pages = {L16},
  year = {2022},
  doi = {10.3847/2041-8213/ac6672}
}

@article{EHTSgrAVI,
  author = {Akiyama, Kazunori and Alberdi, Antxon and Alef, Walter and others},
  title = {{First Sagittarius A* Event Horizon Telescope Results. VI. Testing the Black Hole Metric}},
  journal = {The Astrophysical Journal Letters},
  volume = {930},
  pages = {L17},
  year = {2022},
  doi = {10.3847/2041-8213/ac6756}
}

@article{EHTSgrAVII,
  author = {Akiyama, Kazunori and Alberdi, Antxon and Alef, Walter and others},
  title = {{First Sagittarius A* Event Horizon Telescope Results. VII. Polarization of the Ring}},
  journal = {The Astrophysical Journal Letters},
  volume = {964},
  pages = {L25},
  year = {2024},
  doi = {10.3847/2041-8213/ad2df0}
}

@article{EHTSgrAVIII,
  author = {Akiyama, Kazunori and Alberdi, Antxon and Alef, Walter and others},
  title = {{First Sagittarius A* Event Horizon Telescope Results. VIII. Physical Interpretation of the Polarized Ring}},
  journal = {The Astrophysical Journal Letters},
  volume = {964},
  pages = {L26},
  year = {2024},
  doi = {10.3847/2041-8213/ad2df1}
}

@article{EHTM87PersistentI,
  author = {Akiyama, Kazunori and Alberdi, Antxon and Alef, Walter and others},
  title = {{The persistent shadow of the supermassive black hole of M 87. I. Observations, calibration, imaging, and analysis}},
  journal = {Astronomy \& Astrophysics},
  volume = {681},
  pages = {A79},
  year = {2024},
  doi = {10.1051/0004-6361/202347932}
}

@article{EHTM87PersistentII,
  author = {Akiyama, Kazunori and Albentosa-Ru{\'i}z, Ezequiel and Alberdi, Antxon and others},
  title = {{The persistent shadow of the supermassive black hole of M87. II. Model comparisons and theoretical interpretations}},
  journal = {Astronomy \& Astrophysics},
  volume = {693},
  pages = {A265},
  year = {2025},
  doi = {10.1051/0004-6361/202451296}
}

@article{EHTM87Variability2025,
  author = {Akiyama, Kazunori and Albentosa-Ru{\'i}z, Ezequiel and Alberdi, Antxon and others},
  title = {{Horizon-scale variability of M87* from 2017--2021 EHT observations}},
  journal = {Astronomy \& Astrophysics},
  volume = {704},
  pages = {A91},
  year = {2025},
  doi = {10.1051/0004-6361/202555855}
}

@article{GRAVITY2018,
  author = {Abuter, R. and Amorim, A. and Baub{\"o}ck, M. and others},
  title = {{Detection of orbital motions near the last stable circular orbit of the massive black hole SgrA*}},
  journal = {Astronomy \& Astrophysics},
  volume = {618},
  pages = {L10},
  year = {2018},
  doi = {10.1051/0004-6361/201834294}
}

@article{GRAVITY2023,
  author = {Abuter, R. and Aimar, N. and Amaro Seoane, P. and others},
  title = {{Polarimetry and astrometry of NIR flares as event horizon scale, dynamical probes for the mass of Sgr A*}},
  journal = {Astronomy \& Astrophysics},
  volume = {677},
  pages = {L10},
  year = {2023},
  doi = {10.1051/0004-6361/202347416}
}

@article{Fan2024Fast,
  author = {Fan, Zhong-Ying and Li, Yuehang and Zhou, Fan and Guo, Minyong},
  title = {{Fast magnetic reconnection in Kerr spacetime}},
  journal = {Physical Review D},
  volume = {110},
  pages = {104044},
  year = {2024},
  doi = {10.1103/PhysRevD.110.104044}
}

@article{Fan2025Forces,
  author = {Fan, Zhong-Ying and Zhou, Fan and Li, Yuehang and Guo, Minyong and Chen, Bin},
  title = {{Magnetic reconnection under centrifugal and gravitational electromotive forces}},
  journal = {Physical Review D},
  volume = {111},
  pages = {064067},
  year = {2025},
  doi = {10.1103/PhysRevD.111.064067}
}

@article{Chen2024Plunging,
  author = {Chen, Bin and Hou, Yehui and Li, Junyi and Shen, Ye},
  title = {{Energy extraction from a Kerr black hole via magnetic reconnection within the plunging region}},
  journal = {Physical Review D},
  volume = {110},
  pages = {063003},
  year = {2024},
  doi = {10.1103/PhysRevD.110.063003}
}

@article{Chen2025HotspotPol,
  author = {Chen, Bin and Hou, Yehui and Song, Yu and Zhang, Zhenyu},
  title = {{Polarization patterns of the hot spots plunging into a Kerr black hole}},
  journal = {Physical Review D},
  volume = {111},
  pages = {083045},
  year = {2025},
  doi = {10.1103/PhysRevD.111.083045}
}

@article{Zhang2024Thick,
  author = {Zhang, Zhenyu and Hou, Yehui and Guo, Minyong and Chen, Bin},
  title = {{Imaging thick accretion disks and jets surrounding black holes}},
  journal = {Journal of Cosmology and Astroparticle Physics},
  volume = {05},
  pages = {032},
  year = {2024},
  doi = {10.1088/1475-7516/2024/05/032}
}

@article{Hou2025NearHorizonPol,
  author = {Hou, Yehui and Huang, Jiewei and Guo, Minyong and Mizuno, Yosuke and Chen, Bin},
  title = {{Near-horizon Polarization as a Diagnostic of Black Hole Spacetime}},
  journal = {The Astrophysical Journal Letters},
  volume = {988},
  pages = {L51},
  year = {2025},
  doi = {10.3847/2041-8213/adee09}
}

@article{Zhang2025Autocorrelation,
  author = {Zhang, Zhenyu and Hou, Yehui and Guo, Minyong and Mizuno, Yosuke and Chen, Bin},
  title = {{Autocorrelation signatures in time-resolved black hole flare images: Secondary peaks and convergence structure}},
  journal = {Physical Review D},
  volume = {112},
  pages = {083024},
  year = {2025},
  doi = {10.1103/zmnz-p2rs}
}

@article{Huang2024Coport,
  author = {Huang, Jiewei and Zheng, Liheng and Guo, Minyong and Chen, Bin},
  title = {{Coport: a new public code for polarized radiative transfer in a covariant framework}},
  journal = {Journal of Cosmology and Astroparticle Physics},
  volume = {11},
  pages = {054},
  year = {2024},
  doi = {10.1088/1475-7516/2024/11/054}
}

@article{Zeng2026KerrAdS,
  author = {Zeng, Xiao-Xiong and Wang, Ke},
  title = {{Hotspot image driven by magnetic reconnection in Kerr--anti de Sitter black holes}},
  journal = {The European Physical Journal C},
  volume = {86},
  pages = {272},
  year = {2026},
  doi = {10.1140/epjc/s10052-026-15492-8}
}

@article{Wang2026KerrSen,
  author = {Wang, Ke and Zeng, Xiao-Xiong},
  title = {{Hotspot images driven by magnetic reconnection in Kerr--Sen black hole}},
  journal = {The European Physical Journal C},
  volume = {86},
  pages = {41},
  year = {2026},
  doi = {10.1140/epjc/s10052-025-15257-9}
}

@article{Nathanail2020Plasmoids,
  author = {Nathanail, Antonios and Fromm, Christian M and Porth, Oliver and Olivares, Hector and Younsi, Ziri and Mizuno, Yosuke and Rezzolla, Luciano},
  title = {{Plasmoid formation in global GRMHD simulations and AGN flares}},
  journal = {Monthly Notices of the Royal Astronomical Society},
  volume = {495},
  pages = {1549--1565},
  year = {2020},
  doi = {10.1093/mnras/staa1165}
}

@article{Nathanail2022Reconnection,
  author = {Nathanail, Antonios and Mpisketzis, Vasilis and Porth, Oliver and Fromm, Christian M and Rezzolla, Luciano},
  title = {{Magnetic reconnection and plasmoid formation in three-dimensional accretion flows around black holes}},
  journal = {Monthly Notices of the Royal Astronomical Society},
  volume = {513},
  pages = {4267--4277},
  year = {2022},
  doi = {10.1093/mnras/stac1118}
}

@article{Jiang2023TwoTemperature,
  author = {Jiang, Hong-Xuan and Mizuno, Yosuke and Fromm, Christian M and Nathanail, Antonios},
  title = {{Two-temperature GRMHD simulations of black hole accretion flows with multiple magnetic loops}},
  journal = {Monthly Notices of the Royal Astronomical Society},
  volume = {522},
  pages = {2307--2324},
  year = {2023},
  doi = {10.1093/mnras/stad1106}
}

@article{CruzOsorio2026ElectronTemperature,
  author = {Cruz-Osorio, Alejandro and Meringolo, Claudio and Fromm, Christian M. and Mizuno, Yosuke and Servidio, Sergio and Nathanail, Antonios and Younsi, Ziri and Rezzolla, Luciano},
  title = {{Supermassive Black Hole Imaging with a Self-consistent Electron-temperature Prescription}},
  journal = {The Astrophysical Journal},
  volume = {1001},
  pages = {227},
  year = {2026},
  doi = {10.3847/1538-4357/ae4b3d}
}

@article{Long2026CoolingPolarization,
  author = {Long, Jane SiNan and Uniyal, Akhil and Zhang, Mingyuan and Mizuno, Yosuke and Younsi, Ziri and Fromm, Christian M and Wu, Kinwah},
  title = {{From morphology to variability: radiative cooling effects on horizon-scale polarization in two-temperature GRMHD simulations}},
  journal = {Monthly Notices of the Royal Astronomical Society},
  volume = {550},
  pages = {stag1149},
  year = {2026},
  doi = {10.1093/mnras/stag1149}
}

@article{Liu2002Corona,
  author = {Liu, B. F. and Mineshige, S. and Shibata, K.},
  title = {{A Simple Model for a Magnetic Reconnection--heated Corona}},
  journal = {The Astrophysical Journal Letters},
  volume = {572},
  pages = {L173--L176},
  year = {2002},
  doi = {10.1086/341877}
}

@article{ElMellah2022Corona,
  author = {El Mellah, I. and Cerutti, B. and Crinquand, B. and Parfrey, K.},
  title = {{Spinning black holes magnetically connected to a Keplerian disk - Magnetosphere, reconnection sheet, particle acceleration, and coronal heating}},
  journal = {Astronomy \& Astrophysics},
  volume = {663},
  pages = {A169},
  year = {2022},
  doi = {10.1051/0004-6361/202142847}
}

@article{Gammie2003HARM,
  author = {Gammie, Charles F. and McKinney, Jonathan C. and Toth, Gabor},
  title = {{HARM: A Numerical Scheme for General Relativistic Magnetohydrodynamics}},
  journal = {The Astrophysical Journal},
  volume = {589},
  pages = {444--457},
  year = {2003},
  doi = {10.1086/374594}
}

@article{Moscibrodzka2016Temperature,
  author = {Mo{\'s}cibrodzka, Monika and Falcke, Heino and Shiokawa, Hotaka},
  title = {{General relativistic magnetohydrodynamical simulations of the jet in M 87}},
  journal = {Astronomy \& Astrophysics},
  volume = {586},
  pages = {A38},
  year = {2016},
  doi = {10.1051/0004-6361/201526630}
}

@article{Lyubarsky2005,
  author = {Lyubarsky, Y. E.},
  title = {{On the relativistic magnetic reconnection}},
  journal = {Monthly Notices of the Royal Astronomical Society},
  volume = {358},
  pages = {113--119},
  year = {2005},
  doi = {10.1111/j.1365-2966.2005.08767.x}
}

@article{Broderick2011RIAF,
  author = {Broderick, Avery E. and Fish, Vincent L. and Doeleman, Sheperd S. and Loeb, Abraham},
  title = {{Evidence for Low Black Hole Spin and Physically Motivated Accretion Models from Millimeter-VLBI Observations of Sagittarius A*}},
  journal = {The Astrophysical Journal},
  volume = {735},
  number = {2},
  pages = {110},
  year = {2011},
  doi = {10.1088/0004-637X/735/2/110}
}

\end{document}